\documentclass[useAMS,usenatbib]{mn2e}
\usepackage{amssymb,amsmath,graphicx}
\usepackage{longtable}[1]
\usepackage{graphics}
\usepackage{lscape}
\usepackage{color}
\title[SCs in the Haro 11 starburst]{Super star clusters in Haro 11:  Properties of a very young starburst and evidence for a near-infrared flux excess.}

\author[A. Adamo et al.]{A. Adamo$^{1}$\thanks{E-mail:
adamo@astro.su.se}, G. \"Ostlin$^{1}$, E. Zackrisson$^{1}$, M. Hayes$^{2}$, R. J. Cumming$^{1}$, and G. Micheva$^{1}$\\
$^{1}$Department of Astronomy, Stockholm University, Oscar Klein Center, AlbaNova, Stockholm SE-106 91, Sweden\\
$^{2}$Observatoire Astronomique de l'Universit\'{e} de Gen\`{e}ve, 51, ch des Maillettes, CH-1290 Sauverny, Switzerland}

\newcommand{\araa}{ARA\&A}
\newcommand{\apj}{Ap.J.}
\newcommand{\aj}{A.J.}
\newcommand{\mnras}{MNRAS}
\newcommand{\aap}{A\&A}
\newcommand{\pasp}{PASP}

\newcommand{\msun}{M_{\odot}}

\begin{document}

\date{Accepted xxxx yyyyy nn. Received xxxx yyyyy nn; in original form xxxx yyyyy nn}

\pagerange{\pageref{firstpage}--\pageref{lastpage}} \pubyear{2010}

\maketitle

\label{firstpage}

\begin{abstract}
We have used multi-band imaging to investigate the nature of an
extreme starburst environment in the nearby Lyman-break galaxy
analogue Haro 11 (ESO\,350-IG038) by means of its stellar cluster
population. The central starburst region has been observed in 8
different high resolution {\it HST} wavebands, sampling the stellar
and gas components from UV to NIR. Photometric imaging of the galaxy
was also carried out at 2.16 $\mu$m by NaCo AO instrument at the ESO
Very Large Telescope (VLT). We constructed integrated spectral energy
distributions (SEDs) for about 200 star clusters located in the active
star-forming regions and compared them with single stellar population
models (suitable for physical properties of very young cluster
population) in order to derive ages, masses and extinctions of the
star clusters. The cluster age distribution we recover confirms that
the present starburst has lasted for 40 Myr, and shows a peak of
cluster formation only 3.5 Myr old. With such an extremely young
cluster population, Haro 11 represents a unique opportunity to
investigate the youngest phase of the cluster formation process and
evolution in starburst systems. We looked for possible relations
between cluster ages, extinctions, and masses. Extinction tends to
diminish as function of the cluster age, but the spread is large and
reaches the highest dispersion for clusters in partial embedded phases
($<5$ Myr). A fraction of low-mass (below $10^4 \msun$), very young
($1-3$ Myr) clusters is missing, either because they are embedded in
the parental molecular cloud and heavily extinguished, or because of
blending with neighbouring clusters. The range of the cluster masses is
wide, we observe that more than 30 \% of the clusters have masses above $10^5
\msun$, qualifying them as super star clusters. Almost half of the
cluster sample is affected by flux excesses at wavelengths $> 8000$
\AA\ which cannot be explained by simple stellar evolutionary
models. Fitting SED models over all wavebands leads to systematic
overestimates of cluster ages and incorrect masses for the stellar
population supplying the light in these clusters. We show that the red
excess affects also the {\it HST} F814W filter, which is typically
used to constrain cluster physical properties. The clusters which show
the red excess are younger than 40 Myr; we discuss possible physical
explanations for the phenomenon. Finally, we estimate that Haro 11 has
produced bound clusters at a rate almost a factor of 10 higher than
the massive and regular spirals, like the Milky Way. The present
cluster formation efficiency is $\sim 38$\% of the galactic star
formation rate.
\end{abstract}


\section{Introduction}

Young star clusters (SCs) are luminous and quite compact objects which
in some ways resemble older massive globular clusters. However while
these old systems are "passively" fading due to stellar and secular
evolution, young SCs are still, or in most of the cases, embedded in
their dusty cocoons and at constant risk of destruction by
interactions with the interstellar medium, internal feedback, tidal
forces, and evaporation (see \citealp{2009arXiv0911.0796L} and
references therein). The youngest phases, when the cluster is still
partially embedded in its natal molecular cloud, are difficult to
explore because of the high extinction levels which introduce
uncertainties in age and mass estimates. \citet{2003ARA&A..41...57L}
observed that the embedded phase in young Milky Way (MW) clusters
lasts a few million years. Very young clusters in the Antennae
(\citealp{2002AJ....124.1418W}) system, with ages $\leq$ 4 Myr, show a
wide range of extinction between $0.5 \leq A_V \leq 7.6$
mag. \citealp{2002AJ....124.1418W} identified an additional population
of "red clusters", which they interpreted as young
embedded clusters. Among them, the 2 Myr old WS\,80, with its extreme
$V-I = 2.92$ colour, mass $4 \times 10^6 \msun$ and $A_V = 7.6$ mag, is
the brightest infrared source in the Antennae
system. \citet{2002AJ....124.1418W} indicated roughly 6 Myr as the age
at which a young cluster has gotten rid of the dusty material left over
from its parent cloud. In another galaxy, NGC 4449,
\citet{2008NGC4449R} found 3-5 Myr old clusters with relatively low
extinctions ($ A_V \leq 1.5$ mag). Although these very early stages are
fundamental for our understanding of cluster formation, evolution and
possible destruction (90 \% of clusters are not expected to survive
after their first 10 Myr, \citealp{2003ARA&A..41...57L}), they are
still poorly understood.

The cluster formation process is intimately connected both to the galactic
environment via the star formation efficiency (SFE), and on long time
scales, to the global star formation history (SFH) of the
galaxy. Interactions and/or merging between gas-rich galaxies provide
physical conditions which favour the formation of star clusters and
enhance star formation rates (SFR). Star clusters are easily detected
in active star forming regions inside normal spiral galaxies
(\citealp{2002AJ....124.1393L}; \citealp{2009A&A...503...87C}); in
starburst galaxies, like M\,82 (\citealp{2009ApJ...701.1015K} and
references therein) and NGC\,3310 (\citealp{2003MNRAS.342..259D};
\citealp{2003MNRAS.343.1285D}); interacting
\citep{2005A&A...431..905B} and merging systems
(\citealp{1999AJ....118.1551W}; \citealp{2008MNRAS.384..886V};
\citealp{2006ApJ...641..763W}) to cite only a few
examples. Interestingly, very bright and massive clusters have also
been detected in smaller and less regular systems, like the dwarf
post-starburst galaxies NGC 1569 \citep{2004MNRAS.347...17A} and NGC
1709 (\citealp{2002AJ....123.1454B}; \citealp{2009AJ....138..169A})
and in blue compact galaxies (BCGs), which are active dwarf starburst
galaxies with low metallicity (\citealp{2003A&A...408..887O},
\citealp{adamo09}; Adamo et al.\, in prep.).

The cluster initial mass function (CIMF) and the cluster luminosity
function (CLF) are directly connected to the star formation processes
which act from stellar to galactic scales, due to the clustered nature
of star formation (\citealp{2003ARA&A..41...57L}). In general, the
CIMF can be approximated by a power law relation of the type
$dN(M)/dM=CM^{-\gamma}$, where $\gamma \approx 2$
(\citealp{1999ApJ...527L..81Z}; \citealp{2003A&A...397..473B};
\citealp{2003AJ....126.1836H}, etc.). The CLF, $dN(L) \propto
L^{-\alpha} dL$, on the other hand, has been shown to have wider range
of the index values, $1.8 \leq \alpha \leq 2.7$, and in some cases is
better fit with a double power law with a break in the range $-10.4
\leq M_V \leq -8$ mag (\citealp{2002AJ....124.1393L};
\citealp{1999AJ....118.1551W}; \citealp{2006A&A...446L...9G}). In a
galaxy, the CLF is the synthesis of several populations of SCs
assembled at different ages. For this reason, the CLF contains
imprints of both the mechanisms behind cluster evolution and the
initial environmental conditions where the clusters form. A number of
studies have shown that high star formation levels enable a better
sampling of the LF up to its brighter end and in turn determine the
formation of the brightest cluster in a galaxy
(\citealp{2002AJ....124.1393L}; Gieles et al. 2006a,b;
\citealp{2008MNRAS.390..759B}; \citealp{2009A&A...494..539L};
\citealp{2009MNRAS.394.2113G}). On the other hand, physical factors
also put constraints on the mass (and consequently the luminosity) of
the clusters which are born, such as mass of the parent giant
molecular cloud and the pressure and density of the interstellar
medium (\citealp{2003ARA&A..41...57L} ; \citealp{2002AJ....123.1454B}
, \citealp{2003AJ....126.1836H}). 
\citet{2006A&A...450..129G} noted that if physical conditions in the
galactic environment limit the maximum value of the cluster mass,
leading to a truncated CIMF, this effect will appear in the LF as a
break which corresponds to the maximum cluster mass at the oldest age the galaxy can
form.

In this paper we focus our attention on the blue compact galaxy Haro
11 (ESO\,350-IG038), which has been shown to be extremely efficient in
producing SCs. Generally, BCGs have magnitude between M$_B \approx
-12$ and M$_B \approx -21$ \citep{2000A&ARv..10....1K}. They are
characterized by an ongoing active starburst phase with high star
formation rates (SFRs), producing their typical high surface
brightness and prominent nebular emission line spectra. Photometric
studies of their galactic stellar populations have revealed the
presence of an old underlining stellar component
(\citealp{1988A&A...204...10K}; \citealp{2002A&A...390..891B};
\citealp{genoveva}). Thanks to the {\it HST}'s resolution power, it
has been possible to resolve the BCG burst regions into many bright
star clusters implying that the star formation is remarkably
efficient. However, it still not clear yet how systems with masses
$\leq 10^{10} \msun$ can form such large numbers of star
clusters. \citet{2002AJ....123.1454B}, investigating a sample of 22
nearby ($< 7$ Mpc) dwarf starburst and post-starburst galaxies,
noticed that dwarf starburst galaxies have luminosities brighter than
$M_V=-16$ and pointed out that despite their sizes, they are able to
host very massive clusters. Whether this is a consequence of
stochastic sampling effects of the CIMF in combination with
environmental conditions which favour the formation of star clusters
with $M \geq 10^5 \msun$ or different cluster formation physics is not
yet established. In this paper, we will call clusters with masses
$\geq 10^5 \msun$ super star clusters (SSCs).

In Figure~\ref{h11}, we show a three-colour image of Haro 11. The three
main starburst regions, called knots $A$, $B$, and $C$ by
\citet{2003ApJ...597..263K}, are easily traced in the
heart-shape. Dust filaments cross knot $B$ and are present in all the
burst regions.  Haro 11 is a Ly$\alpha$ and Lyman continuum emitter
\citep{b1} and can be considered a local analog
\citep{2008ApJ...677...37O} of the high redshift Lyman break galaxies
(LBGs). Moreover, Haro 11's IR luminosity of $1.9\times10^{11}$
L$\odot$ (section 6.2) locates Haro 11 in the range of the luminous IR
galaxies (LIRGs). \cite{2001A&A...374..800O}, studying the H$\alpha$
velocity field, found that the mass estimate from rotation curves is
lower than the photometric one, probably because the galaxy is not
rotationally supported, perhaps because the gas is not in dynamical
equilibrium. The total estimated stellar mass is $\approx 10^{10} \msun$. Haro
11's perturbed morphology, multiple-component H$\alpha$ velocity field
and high SFR are all signatures of a merger between a low-mass, evolved
system and a gas-rich component. Interestingly,
\citet{2001A&A...374..800O} have estimated that the present burst in
Haro 11 has been active for $\sim 35$ Myr, and using the mass of neutral hydrogen, H\,{\sc i} 
estimated by \citet{2000A&A...359...41B}, a gas consumption time scale
(that is, remaining lifetime) of $\sim$ 5 Myr for the present burst
phase. We estimated in section \ref{cfr_sfr} that the present SFR in Haro 11 is of $22\pm3 \msun yr^{-1}$. If we take into account the total mass of gas in H\,{\sc i}, H\,{\sc ii}, H$_2$ and photodissociation regions, $M_H^{tot} \sim 2\times10^9 \msun$  \citep{2000A&A...359...41B} we find that the present burst would last at least 100 Myr at its current rate. These time scales show that the gas consumption in the galaxy is
currently very rapid but its future development is highly uncertain.

Finally, the similarity
between this galaxy and the high-redshift LBGs provides an opportunity to
constrain the star-forming mechanisms of the earliest galaxies, when the
metallicity was low and the environments were unrelaxed and chaotic.

This work, together with the analysis done by
\citet{2003A&A...408..887O} and two future publications on the BCGs
ESO\,185-IG03 and Mrk\,930, will be part of a statistical analysis to
constrain and define the major properties of the SSC populations in
BCGs.

The paper is organized as follows. In section 2 and 3 we describe in
detail our data reduction and cluster detection method. Section 4
focuses on the estimation of the clusters' physical properties, in
particular ages, masses and extinction. In this section we also
assess  the evidence for flux excesses at red wavelengths. Section 5
we examine the possible causes of the red excess. In section 6 we
report on the properties of the host galaxy as showed by the cluster
analysis. In section 7 we collect our main conclusions.

The luminosity distance and distance modulus of Haro 11 are estimated to be 82.3 Mpc and (m-M)=34.58 mag\footnote{value taken
from NASA/IPAC Extragalactic Database (NED)}, respectively, assuming a cosmology of
$H_0 = 73$ Km s$^{-1}$ Mpc$^{-1}$, $\Omega_M = 0.27$, and
$\Omega_{\Lambda} = 0.73$ throughout.

\section[]{Data sample and reductions}

A summary of the full data set used in the present
analysis\footnote{Based on observations made with the NASA/ESA
Hubble Space Telescope, obtained at the Space Telescope Science
Institute, which is operated by the Association of Universities for
Research in Astronomy, Inc., under NASA contract NAS 5-26555.} is
given in Table \ref{table-obs}.  We have used archival\footnote{Associated with programs \# GO
9470 (PI D. Kunth) and \# GO 10575 (PI G. \"Ostlin).} {\it HST/ACS}
imaging of Haro 11, with observations from
the FUV to the optical ($\sim 8000 $\AA) wavebands. We have supplemented these data with new \footnote{Associated with programs \# GO 10902, PI:
G. \"Ostlin.} {\it HST/WFPC2}
broadband imaging in V (F606W), I (F814W) and H ({\it NIC 3/F160W}; NICMOS near infrared
NIR). To sample further into the infrared we have used the NaCo adaptive optics (AO) imager in the 
K$_s$ band\footnote{Based on observations made with ESO Telescopes at the
Paranal Observatory under programme IDs 079.B-0585 and 081.B-0234. The observations were carried out the July 24$^{th}$, 2008.}.
Unless otherwise specified, we use the Vega magnitude system
throughout the paper. The zero points in Vega magnitudes are given in Table \ref{table-obs} for all the filters.

\begin{table*}
  \caption{Haro 11: {\it HST} observations. Archival observations are associated with programs \# GO 9470 (PI D. Kunth) and \# GO 10575 (PI G. \"Ostlin). The new observations were obtained in the program \# GO 10902 (PI G. \"Ostlin). The zero point in Vega magnitude system, and the aperture correction, a$_c$ to the point source photometry are included for each filter. The last two columns show the number of objects detected with a $S/N\geq5$, and the corresponding magnitude limit in each frame. (a) Shallower image used only to do photometry in the two central cluster regions saturated in the deeper exposure.}
\centering
  \begin{tabular}{|c|c|c|c|c|c|c|}
  \hline
  Filter &Instrument &  Exposure time & ZP (mag)&a$_c (mag)$ &N($\sigma\leq 0.2$)&  mag limit\\
   \hline
 &Archival obs.&&\\        
   \hline
 F140LP (FUV)&ACS/SBC&2700 s&20.92&-0.49$\pm$0.05&113&25.1\\
 F220W (NUV)&ACS/HRC&1513 s&21.88&-0.41$\pm$0.05&80&23.2\\
 F330W (U)&ACS/HRC&800 s&22.91&-0.35$\pm$0.05&90&23.7\\
 F435W (B)&ACS/WFC&680 s&25.79&-0.50$\pm$0.05&178&27.5\\ 
 F550M (V) &ACS/WFC&471 s&24.88&-0.45$\pm$0.05&144&26.2\\ 
 F814W (I)&ACS/HRC&100 s&24.86&-0.56$\pm$0.07&(a)&(a)\\
 \hline
 &New obs.&&&\\     
\hline    
 F606W (R)&WFPC2/PC&4000 s&22.89&-0.61$\pm$0.09&211&27.5\\
 F814W (I)&WFPC2/PC&5000 s&21.64&-0.73$\pm$0.04&211&26.2\\ 
 F160W (H)&NIC 3&5000 s&21.88&-2.45$\pm$0.37&71&25.6\\
 K$_s$&VLT/NaCo&1222 s&23.32&-2.15$\pm$0.64&66&20.5\\
 \hline
\end{tabular}
\label{table-obs}
\end{table*}

\subsection{{\it ACS} data reduction}
\begin{figure*}
\resizebox{0.8\hsize}{!}{\rotatebox{0}{\includegraphics{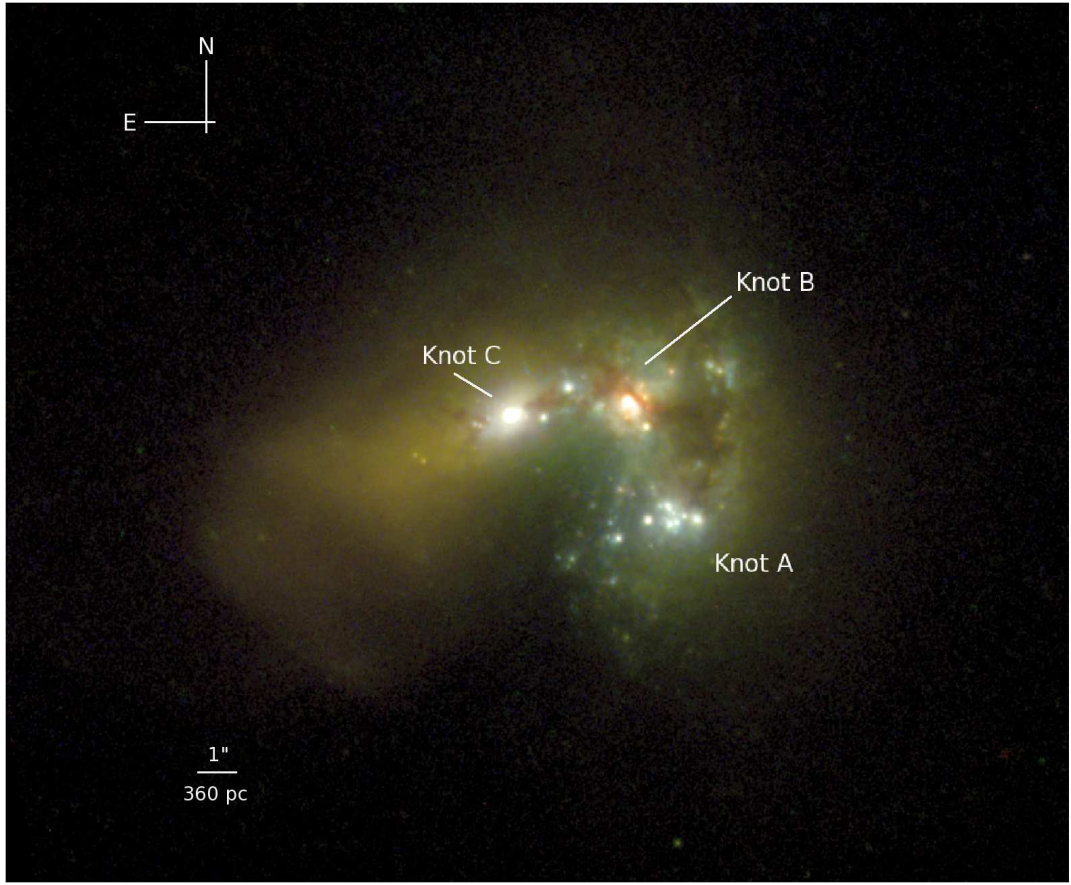}}}
\caption{The starburst regions in Haro 11. The image is the result of
the combination of the {\it WFPC2} F814W filter in red, and the two
{\it ACS} filters: F435W in green and F220W in blue. The orientation
and the $1"$ length are indicated. The names of the three starburst
regions are the same as in \citet{2003ApJ...597..263K}.}
\label{h11}
\end{figure*}

A full description of the data reduction of the archival data is given
in \cite{2009AJ....138..923O}.  The final ACS images have all been
drizzled and aligned using the {\tt MULTIDRIZZLE} task
(\cite{2002hstc.conf..337K}; \cite{2002PASP..114..144F}) in {\tt
PyRAF/STSDAS}\footnote{STSDAS and PyRAF are products of the Space
Telescope Science Institute, which is operated by AURA for NASA} and
rescaled to the HRC pixel scale, that is $0.025 "/px$. The final
science images are rotated to North-up. 
The short {\it HRC} F814W exposure was only used for estimating the flux of the
two brightest sources, in the center of knots $B$ and $C$
(Figure~\ref{h11}). Both knot centers were saturated in the deep {\it
WFPC2} F606W and F814W exposures.

\subsection{{\it WFPC2} data reduction}
Due to the small size (major axes $\sim 13.0"$) of the starburst
regions into the galaxy, the target could easily fit into the
high-resolution (pixel scale is $0.045"$) Planetary Camera (PC) which
has a FOV of $35"\times35"$. In order to refine the WFPC2 resolution
we used sub-pixel dithering and combining patterns. The data reduction
were performed with {\tt MULTIDRIZZLE}. 
We took into account sky corrections during the photometric analysis rather than performing
sky-subtraction with {\tt MULTIDRIZZLE}. Before combining the final
image, we resized the pixel scale to $0.025"$, applying a drop size of
$0.8$ for both the filters, and rotated the final image so that North was up.

\subsubsection{The {\it WFPC2} point spread function}

Because we found no isolated bright objects either in the galaxy or in
the PC field, we constructed a filter-dependent point spread function
(PSF) using archival data. We retrieved {\it WFPC2}/F814W images of
the NGC-104-F field of the globular cluster 47 Tuc (\# GO 8160, PI I.\
King). This field was chosen because it is not crowded and has
dithering path and exposure time comparable to our data.  We built the
F606W PSF in a similar fashione using data from another
uncrowded stellar field (\# GO 8090, PI S. Casertano). We reduced the
images applying the same method, taking into account only the PC field
of the mosaic. Isolated sources in the field were selected by eye, and
encircled energy distributions for each of them were inspected in
order to avoid sources with irregular PSF shapes. 
The tasks {\tt PSTSELECT} and {\tt PSF} were
used to produce the final PSF. These two constructed PSFs were used to
perform a completeness test (see section 3.2).

\subsection{{\it NICMOS} data reduction}  

The NIC3 camera was chosen because of the size of the field of view,
$51.2" \times 51.2"$. To sub-sample and improve the pixel scale
resolution (for NIC3, the plate scale is $\sim 0.2"/px$) a double
NIC-SPIRAL-DITH path with 3 dithered pointings each was applied in
order. The single science frames showed significant differences in the
sky background levels and presence of bias residuals, which we
corrected before running {\tt MULTIDRIZZLE}. In order to get the best
combination of pixel sub-sampling and pixel drop size values we
compared the flux growth curves of some bright point-like objects in
the final science frame with a reference model PSF produced by {\tt
TinyTim}
software\footnote{http://www.stsci.edu/software/tinytim/}. The
resulting best set of input values for the final combined image was
found to be {\tt pixscale} $= 0.067"/px$ and {\tt pixfrac} $=0.7$. The
pixel scale applied to the final frame is the diffraction limited
value for the NIC3 camera. The final science frame was rotated to
North-up and calibrated in units of $DN/s$. 
We performed a
number of tests of the calibration of the final IR frame: these are
detailed in Appendix A.

\subsection{NaCo data reduction}

Ks imaging was carried out at the VLT UT4 telescope using Naos-Conica
(NaCo) AO instrument in natural guide star mode. The CONICA imager and
spectrograph equipped with the Aladdin3 array and S27 camera (FOV of
$28"\times28"$) was chosen. The chopping mode and
dithering path were set and a total of 24 target images and 24 sky
fields were taken, each with $DIT = 9.4 s$ averaged on 13 $NDITs$. We used {\tt PyPELINE} \citep{genoveva} for bad-pixel
masking, pair subtraction, flat field correction, source masking, sky
background subtraction and alignment. Before the final stacking, the
quality of every frame was checked and those with blurred PSFs were
excluded, in order to achieve the best possible resolution. Frames
with PSF full width at half maximum (FWHM) $> 4 px$($\sim 0.108"$) were rejected. The final image was built from the median of
the best 10 selected frames (corresponding to a total exposure time of 1222 s). A standard star was observed the same
night, with the same instrumental set, and at comparable airmass
values that for our target (the difference is $< 0.08$) so that the ZP
value is affected by the same atmospheric extinction value as the
galaxy. The pixel scale in the final frame was $0.027"/px$.

\section{Cluster selection and photometry}
\subsection{Cluster detection} \label{cluster-det}
{\tt SExtractor} \citep{1996A&AS..117..393B} was used to find sources in the deep
R and I {\it WFPC2} exposures. The starburst in the galaxy is intense
and many point-like sources lie in crowded regions and superimposed on
the bright galactic background. Farther out of the main body,
point-like sources can be still observed but now in empty regions with
lower background values. 
In order to extract candidate clusters in all
these different regions of the galaxy field we ran {\tt SExtractor}
twice on each frame, each time with different input parameters. To
find sources in the outskirts of the galaxy, we used a detection
threshold of $2\sigma$ above the background RMS noise, a minimum of 4
adjacent pixels and a minimum contrast for de-blending set to the
default value, 0.001.  To detect sources into the starburst region we
increased the detection threshold to $2.5 \sigma$, applied one of the
mexhat filters available in {\tt SExtractor}, and raised the value of
the de-blending minimum contrast to 0.005. In this way we avoided
large numbers of spurious detections while at the same time being able
to separate blended sources in crowded regions. The two lists of
coordinates were then merged into a unique catalogue, cleaned from
multiple detections of the same object and from sources separated by
less than the diffraction limit of the telescope. We fixed a center
for the galaxy ($\alpha = 00:36:52:62.7$, $\delta=-33:33:17.49$) and
delimited detections to a region of radius of $13.75"$ around the
center. Finally, detections in the two cleaned R and I band catalogues
were cross-checked. All the sources which were not detected in both R
and I catalogues were discarded. The final catalogue comprised 563
objects. 

For the two {\it WFPC2} images, we performed a correction for CTE
(charge transfer efficiency) effects on the observed fluxes using the
algorithm of \cite{2009PASP..121..655D}. Because of the complexity of
the region, i.e. point like sources located in the extended luminous
background of the galaxy, we chose not to use the median background
value, as prescribed by Dolphin's formula.  Instead we estimated a
"local'' background for each source, given by the median value of the
pixel columns above the source, where few a percent of the total
charge has probably been trapped. 

\begin{figure}
\resizebox{\hsize}{!}{\rotatebox{0}{\includegraphics{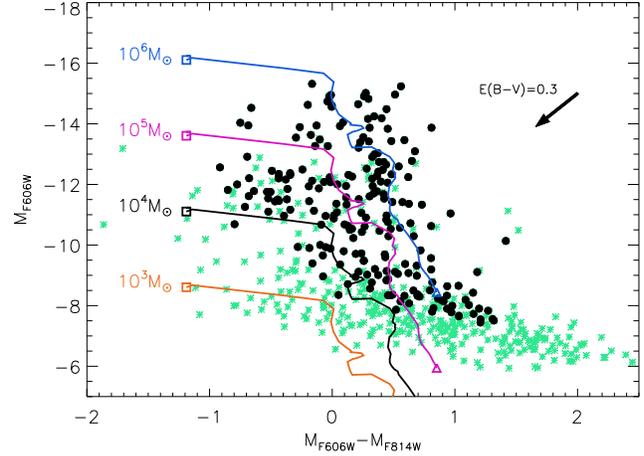}}}
\caption{Colour-magnitude diagram for the 563 cluster candidates
detected in both R and I filters. The black filled dots are the 211
sources with a $S/N \geq 5$ in both filters (see Table
\ref{table-obs}). Model evolutionary tracks with metallicity $Z=0.004$ are from \citet{Zackrisson
et al. a}. The masses corresponding to each track are indicated. The thick squares show the model starting point at 1 Myr, the triangles the end at 14 Gyr. The
arrow shows the direction of a correction for extinction of
$E(B-V)=0.3$.}
\label{CMD}
\end{figure}

We show the 563 cluster candidates in the colour-magnitude diagram
(CMD) in Figure~\ref{CMD}.

\subsection{Completeness limits}
\label{completeness}

\begin{figure}
\resizebox{\hsize}{!}{\rotatebox{0}{\includegraphics{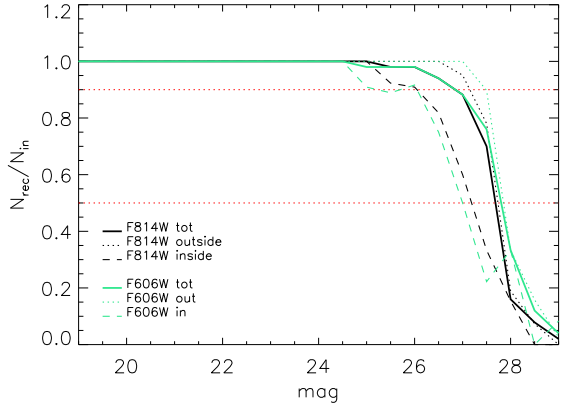}}}
\caption{Recovered fractions of sources as function of their magnitude
for R and I images. The two continuous thick lines are the
completeness limits for detection in filters F814W (black) and F606W
(green). The dotted (dashed) line show the recovered fractions inside
(outside) the crowded starburst region, as defined in section
\ref{cluster-det}. The two red dotted horizontal lines show the $90\%$ and
$50\%$ completeness levels.}
\label{compl}
\end{figure}

To test the method we used to constrain the source catalogue and the
detection limits, we performed a completeness analysis for both R and
I {\it WFPC2} frames. A mock catalogue was made with objects
positioned along a grid of $30$ pixels per side (big enough not to
increase crowding). All the positions were located inside the circle
defined previusly. Magnitudes were randomly assigned to each position,
from $19.0$ mag to $29.0$ mag with an interval of 0.5. For each
filter, we created a synthetic frame with the sources using {\bf
mksynth}, a task of the in the public software toolkit {\bf Baolab}
\citep{1999A&AS..139..393L}. We supplied the PSF shape. 
The object detection was done in the same way as previously explained and no conditions were placed on the photometry. We then
counted how many of the input objects were recovered, allowing a
displacement not larger than 1.5 pixels. We repeated the procedure
three times, shifting the grid by $10\times10$ pixels and re-assigning
magnitude randomly to prevent introduction of systematics. The mean
recovered fractions of objects as function of magnitude is shown in
Figure~\ref{compl}. The completeness goes down to 90 \% at mag 27 in
both filters, and only 50\% of objects are recovered at $\sim 28.0 $
mag. 

This is an averaged result: in reality, incompleteness due to crowding
in the central active star-forming region is larger than in the
outskirts of the galaxy. To assess by how much, we did a second
completeness estimation, this time distinguishing between the
two regions, with boundary radius fixed at $6.5"$ from the center. The
fraction of recovered objects inside the crowded region (dashed and
dotted lines in Figure~\ref{compl}) begins to diminish in both R and I
frames at between $25.0 $ mag and $25.5 $ mag, reaching $90\%$ already
at $\sim 26.0$ mag and dropping to $70\%$ at $27.0$ mag. On the other
hand, the recovered fraction of objects outside the crowded region gives
$100 \%$ completeness for detections as faint as $27.0 $ mag. As expected,
blending affects the detections at the faint end of the cluster
luminosity function in the active starbursting region. 

\subsection{Photometry and final catalogue}
\label{photometry}
Aperture photometry was performed with the task {\tt PHOT}, in all the
available filters, using the catalogue previously determined. The Iraf
tasks {\tt GEOMAP} and {\tt GEOXYTRAN} were used to transform the
coordinates from the {\it WFPC2} system to the corresponding {\it
ACS}, {\it NICMOS}, and NaCo/S27 catalogues. Growth curves of several
objects in different filters were investigated to decide the size of
the aperture radius. Finally, we carried out photometry in all frames
using a fixed aperture radius of $a_r = 0.1"$ and estimating the local
sky background in an annulus of internal radius $s_r =0.125"$ and
width $0.05"$. In order to to avoid photometric uncertainties
affecting the age and extinction estimates, we rejected for each
filter data points with magnitude errors $\sigma_m > 0.2$. In Table \ref{table-obs} we show the corresponding magnitude limits and the number of objects in each frame after this selection. For the
final catalogue, we retained only objects detected in at least 3
filters. In total it contains 198 objects\footnote{The complete
photometric catalogue is available at
http://www.astro.su.se/$\sim$adamo or on request from the
authors.}. Galactic extinction corrections in all the filters was
applied. Only 17\% (34/198) of sources had detection in 3 filters, 21\% in 4, 12\% in 5, 11\% in 6, 13\% in 7, 12\% in 8, and 14\% (28/198) in the 9 available filters. In total, 83\% of the objects in the catalogue had detections in at least 4 filters.

Aperture corrections and associated errors were estimated for each filter frame and added to the fluxes and photometric errors which we previously determined. They are given in Table \ref{table-obs}. We used
{\tt TinyTim} PSF simulator to estimate the aperture corrections
values for {\it ACS} and {\it WFPC2} frames. A different method was
applied to the two IR filter frames, H and K$_s$. The H frame has a
wider PSF and lower pixel scale resolution. For these reasons, we
decided to estimate the aperture correction directly from isolated
bright star-like objects in the frame. In the large FOV of the $NIC3$
camera we could identify some point like objects. Moreover an
identical set of {\it NICMOS} data was available for other two BCGs,
Mrk 930 and ESO\,185-13 (part of the same project \# GO10902, see
\cite{adamo09}; results will be published in a following up paper,
Adamo et al., in prep.) and also in those frames, reduced in the same
way, few star-like objects were available.  The aperture used to do
photometry ($0.1"$) was quite small and corresponded to half a size of
the original {\it NIC3} pixel scale ($0.2"$), making the error due to
the aperture correction large. On the other hand, using bigger radius
apertures would introduce blending of sources. 

The aperture correction for the K$_s$ image was derived from 3 bright
isolated point-like objects in the frame. In this case the problem was
not the poor pixel resolution scale, but the effect of the AO
technique on the PSF shape. The flux of the source is split between
the diffraction-limited gaussian PSF core and the extended wings,
making the aperture correction PSF-dependent. As expected, the 
aperture correction value is high $a_c=-2.15 \pm 0.64$, because
a considerable part of the flux is contained in the PSF wings and the
associated error is large.

\section{SED fitting and cluster analysis}


\subsection{Models}

For our SED fits of the target clusters, we have used the
\citet[][hereafter Z01]{Zackrisson et al. a} spectral synthesis
model. This model predicts the combined SED of both stars and
photoionized gas. Often neglected, the gas component can
actually have a pronounced impact on the broadband fluxes of very
young stellar populations \citep[e.g.][]{Krüger et al.,Anders &
Fritze-Alvensleben,Zackrisson et al. b}.

The Z01 model, which includes pre-main sequence evolution and a
stochastic treatment of horizontal branch morphologies at low
metallicities, is based on synthetic stellar atmospheres by
\citet{Lejeune et al.} and \citet{Clegg & Middlemass}, together with
stellar evolutionary tracks mainly from the Geneva group. The gas
continuum and emission lines are predicted using the photoionisation
code Cloudy, version 90.05 \citep{Ferland et al.}. At each time step,
the spectral energy distribution of the integrated stellar population
is fed into Cloudy, which then provides a realistic evolution of the
emission-line ratios over time. The spectra
predicted by the models are redshifted to match the target galaxies
and convolved with the transmission profiles of the filters used.

In the Z01 model, we assume
a stellar metallicity of $Z_\mathrm{stars}=0.004$, a Salpeter initial
mass function ($\mathrm{d}N/\mathrm{d}M\propto M^{-2.35}$) throughout
the mass range 0.08--120 $M_\odot$ and an instantaneous burst (i.e.\
single-age) population. A conversion factor is applied to correct the
estimated cluster masses to a Kroupa IMF \citep{2001MNRAS.322..231K}
in the mass range 0.01--120 $\msun$. The properties of the nebular
component are determined by the adopted gaseous metallicity
$Z_\mathrm{gas}$, the hydrogen number density $n(\mathrm{H})$, the gas
filling factor $f$ and the gas mass $M_\mathrm{gas}$ available for
star formation. We adopt $n(\mathrm{H})=100$ cm$^{-3}$ and $f=0.01$
(values typical for H\,{\sc ii} regions), a gas metallicity identical
to the stellar one ($Z_\mathrm{gas}=Z_\mathrm{stars}$) and a gas mass
of $M_\mathrm{gas}=10^6\ M_\odot$. Finally, we have used the spectral
synthesis model of \citet[][hereafter M08]{Marigo et al.}  to test the robustness of
some of our conclusions. This model features a more sophisticated
treatment of thermally pulsating asymptotic giant branch (AGB) stars
(which mainly affects the interpretation of near-infrared data), but
does not include nebular emission. For this reason it is not suitable
for the analysis of clusters at ages $\leq 10$ Myr \citep{Zackrisson
et al. a}.

\subsection{Constraining the physical properties of the SCs}
\subsubsection{Chi square method} 

We used a least-squares fit to estimate for each cluster the best fit
to our models (similar to other previous works, see \citealp{2004MNRAS.347..196A}; \citealp{2003A&A...397..473B}). 
Internal extinction was treated as a free
parameter, and we used the Calzetti extinction law
\citep{2000ApJ...533..682C}, allowing extinction varying from
$E(B-V)=0.0$ to 3.0 with a step of 0.01. The {\it
Q-value}\footnote{also known as the $\chi^2$ probability function, see
\citet{1992nrfa.book.....P}. Best-fit models with values of $Q$ which
exceed 0.1, are good representative of the observed data; values of $Q
> 0.001$ can be considered still acceptable; while models with lower
values are unlikely.} and the reduced $\chi^2_r=\chi^2/\nu$ were calculated. The $\nu=N-m$ is the number the degrees
of freedom. In our case, for each cluster, the number of detections in
different filters, $N \geq 3$, while the number of parameters $m=3$
(the age, mass and extinction of the cluster). The
{\it Q-value} is a statistical measurement of the goodness of the fit
\citep{1992nrfa.book.....P}. $Q$ and $\chi^2_r$ were estimated only for
clusters with at least detection in 4 different filters ($N>3$).

Among our initial model fits, we noticed several cases for which
$\chi^2_r >>1$ and $Q < 10^{-5}$. For those cases, the inspection of
the residuals (displacements between the modeled and the observed
integrated photometry of the cluster, $\Delta m =
m_{\textnormal{mod}}-m_{\textnormal{obs}}$), revealed an unusual
positive offset, or "excess", of the observed fluxes in H, K$_s$, and
I bands.  

We performed two further sets of $\chi^2$ fits, this time excluding
the NIR data. The first of these we refer to as {\it UV-UBVR}, where
only the 6 bluer filters were included in the fit.  The second set included 7 filters,
FUV, NUV, U, B, V, R, and I, indicated as {\it UV-UBVRI}.  The initial
fit, including all the available data from UV to near-IR, we denote
{\it UV-UBVRIHK$_s$}. We maintained the requirement of at least
detections in 3 filters for each SED fit. This restriction meant that
we could fit 190 objects when the H and K$_s$ bands were
excluded in the {\it UV-UBVRI} fit. Of these, 35\% have
detections in all the 7 included bands, while 22\% have detections in
only 3 filters.  When the I band was excluded, the number of objects
fitted decreased to 149 (23\% with at least 3 detections and 45\% with
detections in all the allowed filters).

\subsubsection{Investigating the excess at near-IR wavelengths}
\label{red-excess}
\begin{figure*}
\resizebox{0.48\hsize}{!}{\rotatebox{0}{\includegraphics{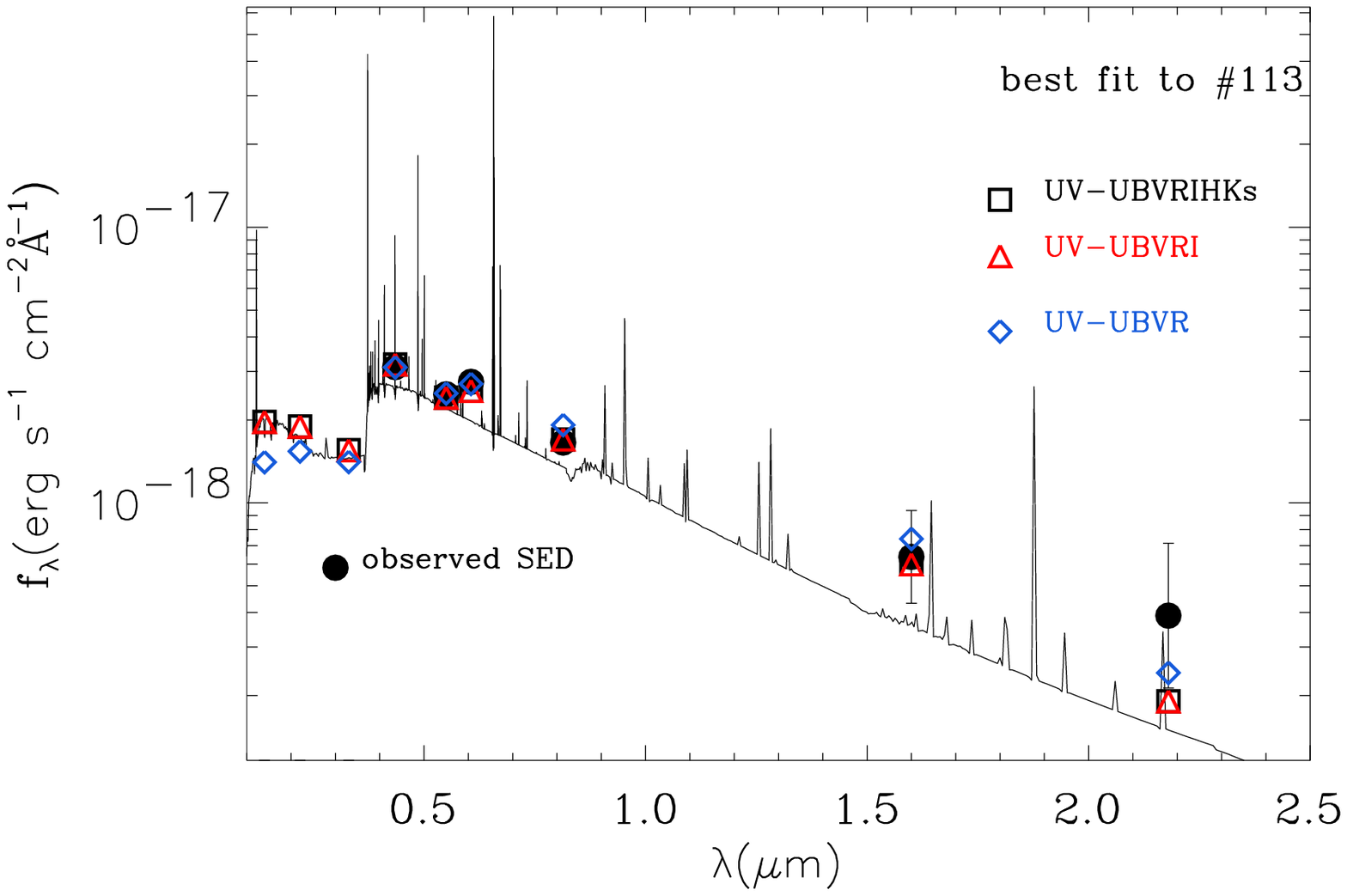}}}
\resizebox{0.48\hsize}{!}{\rotatebox{0}{\includegraphics{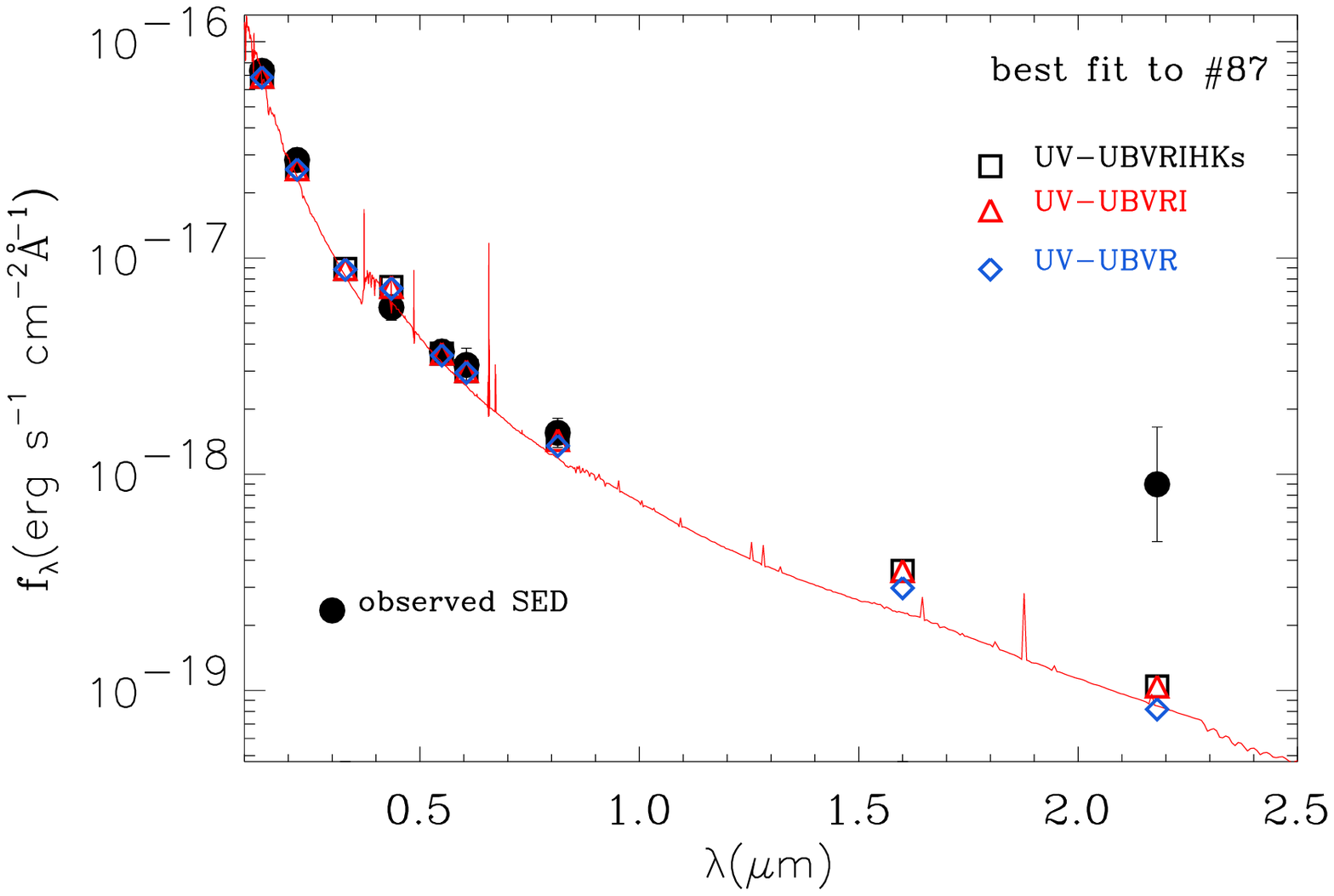}}} \\

\resizebox{0.48\hsize}{!}{\rotatebox{0}{\includegraphics{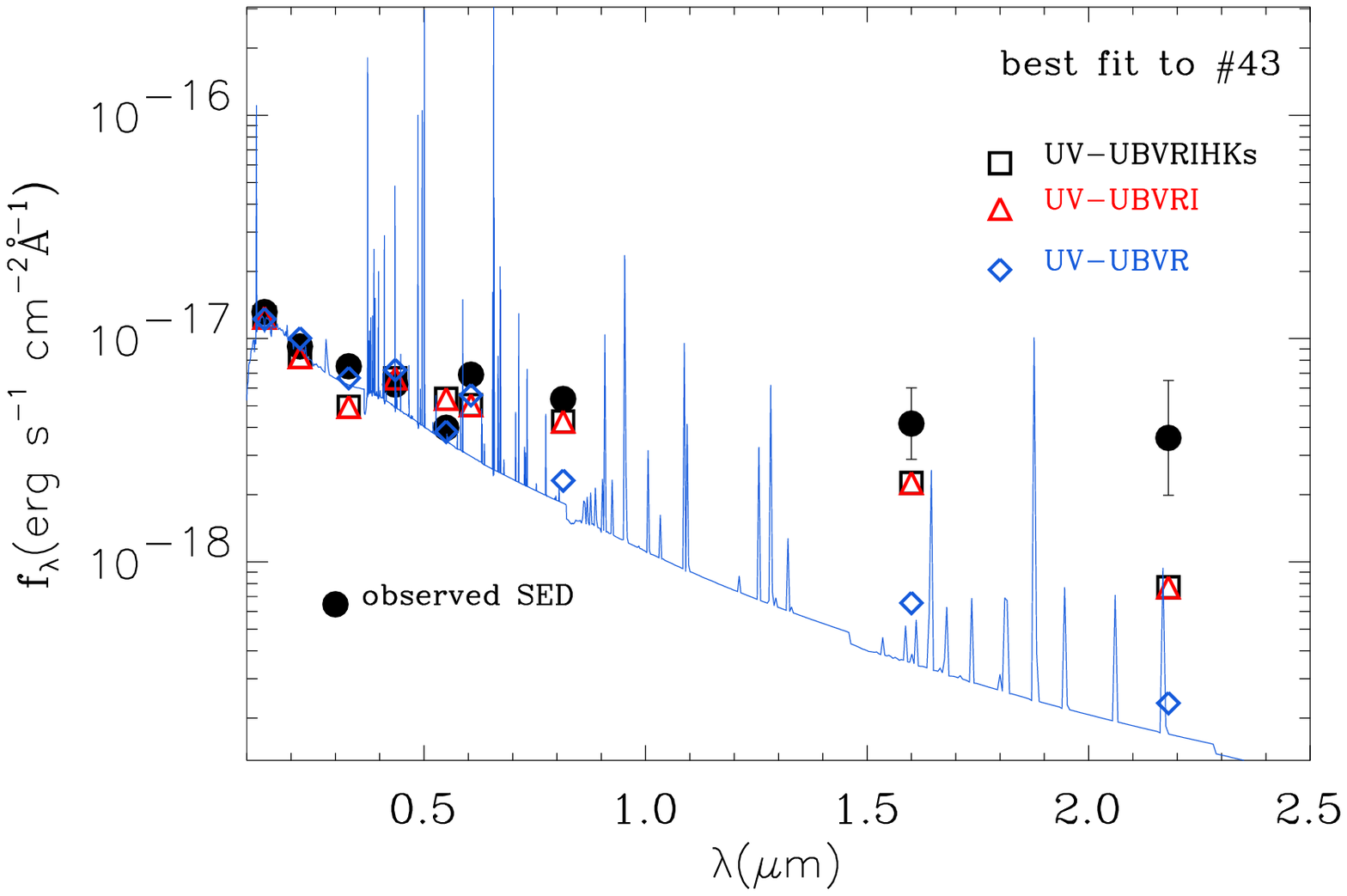}}}
\resizebox{0.48\hsize}{!}{\rotatebox{0}{\includegraphics{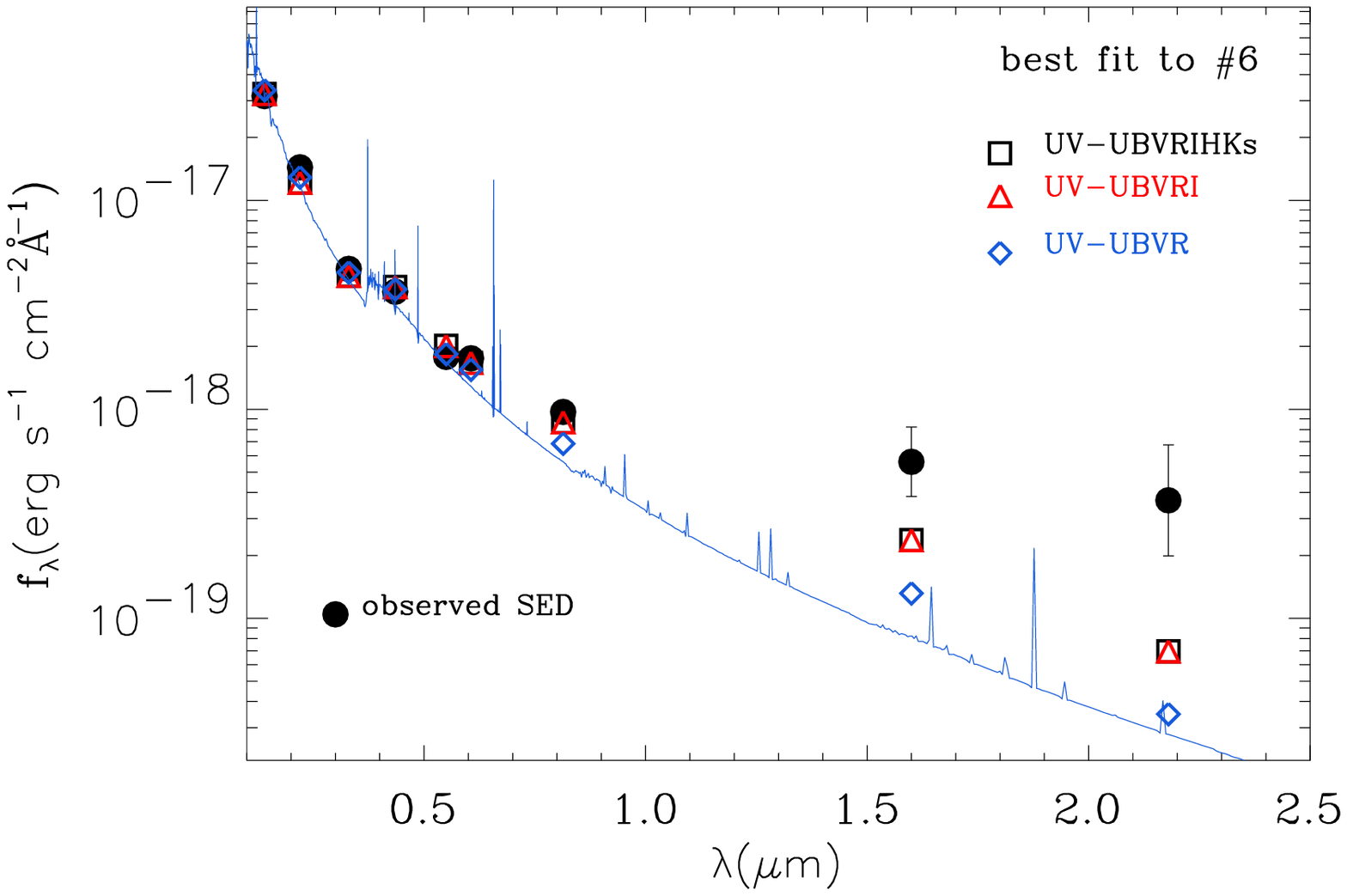}}} \\
\caption{SED analysis of 4 clusters. The filled black points indicated
the observed photometric values of each cluster. The integrated model
fluxes are labelled with different symbols for each set of fit as
indicated in the plots. We plotted the spectrum of the final best
fitting model with a colour which corresponds to the type of fit.}
\label{spec}
\end{figure*}

In this section we present "observational" evidence for the red excess and at the same time probe that the fit to the UV-blue optical integrated fluxes of the clusters produces robust constraints of the cluster properties.

In Figure~\ref{spec} we show an example of the three different fits to
the integrated SED data points for four different clusters with
different physical properties (a table with integrated photometry of
these clusters is presented in Appendix B). The quantities plotted
have been converted from Vega magnitudes to physical flux units,
erg~s$^{-1}$cm$^{-2}$\AA$^{-1}$.  Integrated photometric data points
with associated error bars are plotted together with the best
synthetic data points obtained by the three different types of fit:
{\it UV-UBVRIHK$_s$} (black squares), {\it UV-UBVRI} (red triangles)
and {\it UV-UBVR} (blue diamonds). The underlying spectrum represent
the best model. The broad F606W filter
transmits the prominent H$\alpha$ emission line. So in cases where we
see a pronounced jump between fluxes in the narrow F550M and the broad
F606W filter we expect young ($\leq 8.5$ Myr) stellar complexes
(e.g. see clusters \#43 and \#113 in Figure~\ref{spec}). The
displacement between these two data points is negligible already
around 10 Myr (e.g.\ cluster \#6 in Figure~\ref{spec}). Finally, the
data points sampling the U-UV part of the spectrum are extremely
sensitive to the internal extinction of the cluster. In Table
\ref{fitstats} we list the statistical results and best fit parameters
obtained with all three sets of fits.

\begin{table}
  \caption{Final outputs given by the three sets of SED fits as shown in Figure~\ref{spec}. In bold we show the final age, mass and extinction assigned to the clusters.}
\centering
  \begin{tabular}{|c|c|c|c|c|c|}
  \hline
  \hline
 &\multicolumn{4}{|c|}{{\it UV-UBVRIHK$_s$}} \\    
 id&$\chi_r^2$& $Q$&Myr&$10^5 \msun$&E(B-V)\\
  \hline
 {\bf113}& {\bf0.7}& {\bf0.72}& {\bf5.5}& {\bf1.4}& {\bf0.44}\\
 87&1.65&0.25&10.5&1.3&0.0\\
 43&13.6&$<10^{-5}$&45.0&16.3&0.23\\
 6&3.28&0.006&12.5&0.8&0.0\\
 \hline
 \hline
 & \multicolumn{4}{|c|}{{\it UV-UBVRI}}\\  
 \hline
 id&$\chi_r^2$& $Q$&Myr&$10^5 \msun$&E(B-V)\\
  \hline 
 113&0.45&0.5&5.5&1.4&0.44\\
 {\bf87}& {\bf1.1}& {\bf0.34}& {\bf10.5}& {\bf1.3}& {\bf0.0}\\
  43&13.1&$<10^{-5}$&45&16.0&0.23\\
   6&2.6&0.008&12.5&0.8&0.0\\
   \hline
 \hline
 &\multicolumn{4}{|c|}{{\it UV-UBVR}} \\  
  \hline
 id&$\chi_r^2$& $Q$&Myr&$10^5 \msun$&E(B-V)\\
  \hline
 113&-&-&5.5&1.8&0.5\\
 87&0.8&0.31&9.5&1.2&0.0\\
 {\bf43}& {\bf1.4}& {\bf0.07}& {\bf3.5}& {\bf1.5}& {\bf0.34}\\
  {\bf6}& {\bf1.3}& {\bf0.41}&{\bf8.5}& {\bf0.5}& {\bf0.01}\\
 \hline
\end{tabular}
\label{fitstats}
\end{table}

The fit to cluster \# 113 is quite good all the way from the UV to the
IR (inside the error bars). Looking at the Table~\ref{fitstats}, we see that removing filters
from the fit produces no noticeable effect. The jump between V
and R clearly confirms the young age of this object. The non-detection in the UV part of the spectrum is in agreement with the high extinction value. The properties of
83 clusters with similar SED fits to this cluster were estimated by the {\it
UV-UBVRIHK$_s$} fit.

The {\it UV-UBVRIHK$_s$} fit to the cluster \#87 clearly failed to fit
the integrated IR flux in K$_s$, sitting above the best-fit model with
an excess of $\Delta m_{Ks} = 2.34$ mag. The large error associated
with the observed flux in K$_s$ produces a smaller effect on the
$\chi_r^2$ and $Q$ values of the {\it UV-UBVRIHK$_s$} fit. Taking into
account this factor, even small differences in $\chi_r^2$ and $Q$, as
observed here between the {\it UV-UBVRIHK$_s$} and {\it UV-UBVRI}
fits, become important. In this case, we discarded the {\it UV-UBVR}
fit solution because, as indicated in Table~\ref{fitstats}, the
exclusion of the I band from the fit does not give any improvement. We
found 26 clusters with an excess at the IR wavelengths (H and K$_s$). For these objects, the final ages, masses and extinctions were derived from a {\it
UV-UBVRI} fit.

The fits to cluster \#43 give even more dramatic results. Neither the {\it
UV-UBVRIHK$_s$} nor the {\it UV-UBVRI} fits are good (high $\chi_r^2$
and high $Q$). The estimated age, 45 Myr, is impossible to reconcile
with the jump observed between the narrow V and broad R filters, and
the estimated mass is high. Excluding the I band photometry gave a
better fit (see statistical values in the Table 2). The {\it UV-UBVR}
fit to this cluster produces acceptable $\chi_r^2$ and $Q$, indicating
a cluster only 3.5 Myr old, with mass $\sim 10^5 \msun$. E(B-V) is
0.34, which corresponds to a visual extinction of $A_V=1.24$. The
observed excesses compared to the best {\it UV-UBVR} model are 0.91,
2.01 and 2.97 mag in I, H and K$_s$, respectively.  A similar case is
cluster \#6. Here, too, the flux excess is not confined to the NIR
part of the spectrum, but also extends into the I-band, if less
dramatically than \#43. In \#6 the V-R jump is also less pronounced,
in agreement with its best estimated age, 8.5 Myr. We found a total of
78 objects affected by flux excess at wavelengths $> 8000$ \AA\ like
clusters \#6 and \#43. For these clusters, we have determined
physical properties from the {\it UV-UBVR} fit.

After this first look, it is evident that while our SSP models can
successfully model the optical and UV portions of the observed cluster
SEDs, they are unable to reproduce the NIR photometry of many of the
clusters. This suggests that the models we have used are missing a
source of emission which is bright in the near-infrared.

\begin{figure}
\resizebox{\hsize}{!}{\rotatebox{0}{\includegraphics{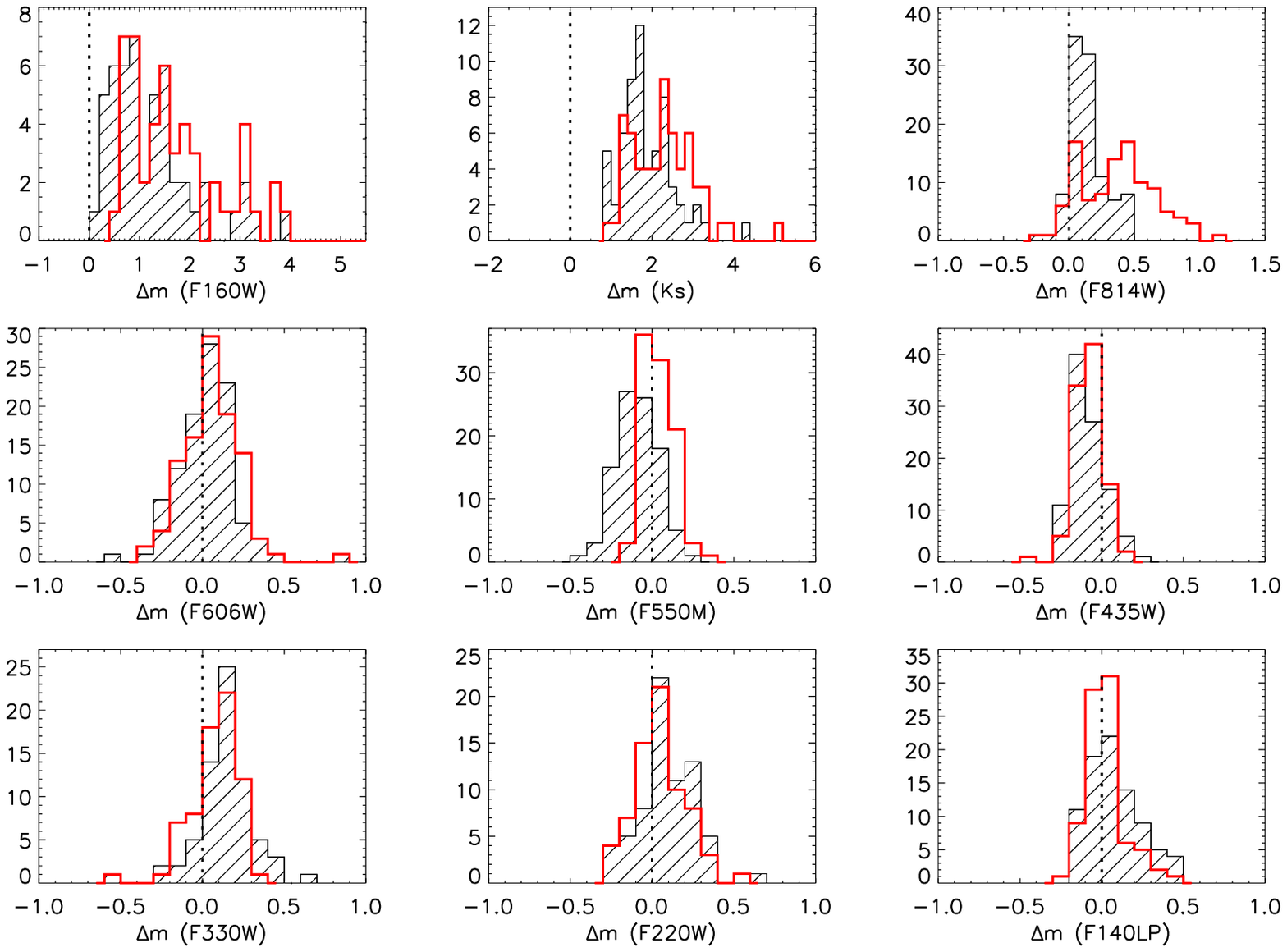}}}
\caption{Residuals ($\Delta m = m_{\textnormal{mod}}-m_{\textnormal{obs}}$) 
of the $\chi^2$ SED fitting in all the available filters for sources with
detected excess in H.  The hatched black histograms show all these
clusters; the histograms with red borders are the residual
distribution for the same targets but where I (if the flux excess is present), H and K$_s$ have been
excluded from the SED fitting. The vertical thick dotted lines
indicates zero residuals, i.e.\ a match between model and observation.
The number of objects differs between plots because of the different
detection thresholds in each band. }
\label{delta}
\end{figure}

In Figure~\ref{delta} we show how the three different fit types affect
the residuals in the SED fits in all filters, for targets with flux
excess in the H band. The black hatched histograms are the residuals
from the {\it UV-UBVRIHK$_s$} fit relative to the observed data
points, $\Delta m = m_{\textnormal{mod}}-m_{\textnormal{obs}}$, for
all the clusters. The residuals in H and K$_s$ are systematically
offset toward positive $\Delta m$.  In all the other filters the
spreads are large, but the residuals are evenly distributed around
zero. When the {\it UV-UBVRI} and {\it UV-UBVR} fits were performed,
the spread in the residuals of the excluded filters (red thick histograms for F160W, K$_s$, and F814W ) were even larger. 
On the other hand, the improvements in the distributions for R and blue-ward filters
are clear. The red histogram of the F814W filter has a secondary peak around zero. This peak is produced by the clusters which do not have any flux excess in I band and for this reason fitted by the {\it UV-UBVRI} fit.

\begin{figure*}

\resizebox{0.45\hsize}{!}{\rotatebox{0}{\includegraphics{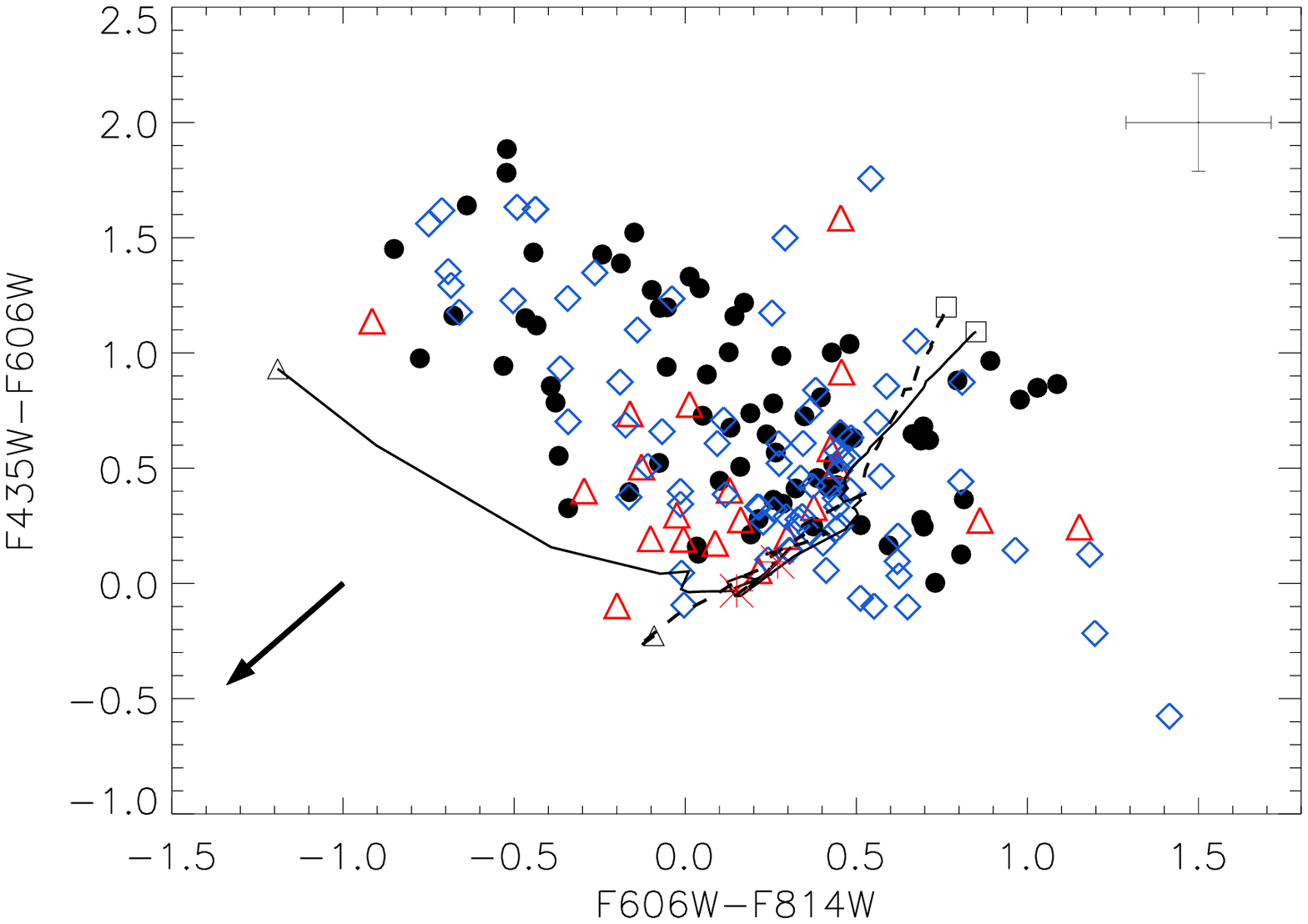}}}
\resizebox{0.45\hsize}{!}{\rotatebox{0}{\includegraphics{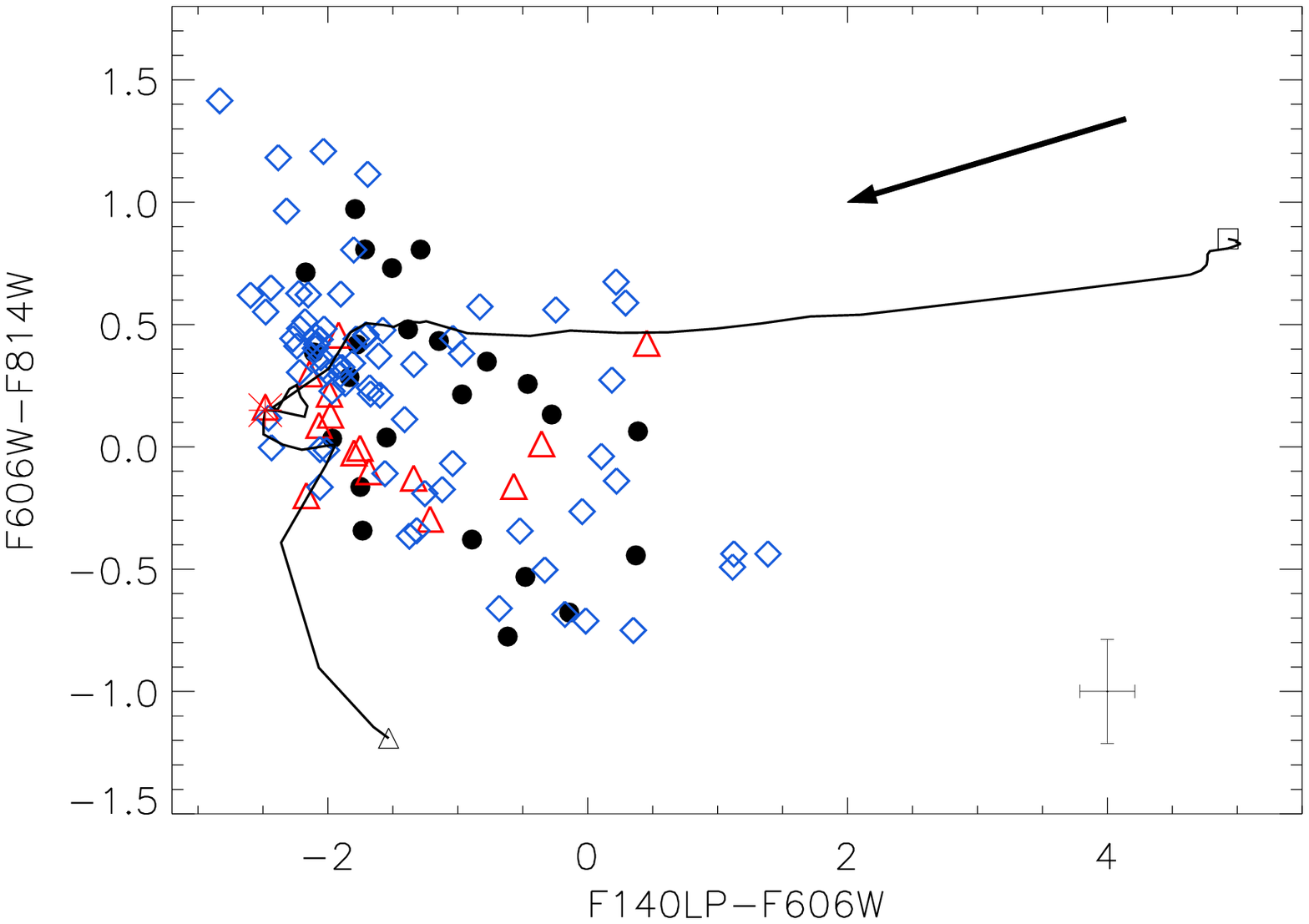}}} 
\resizebox{0.45\hsize}{!}{\rotatebox{0}{\includegraphics{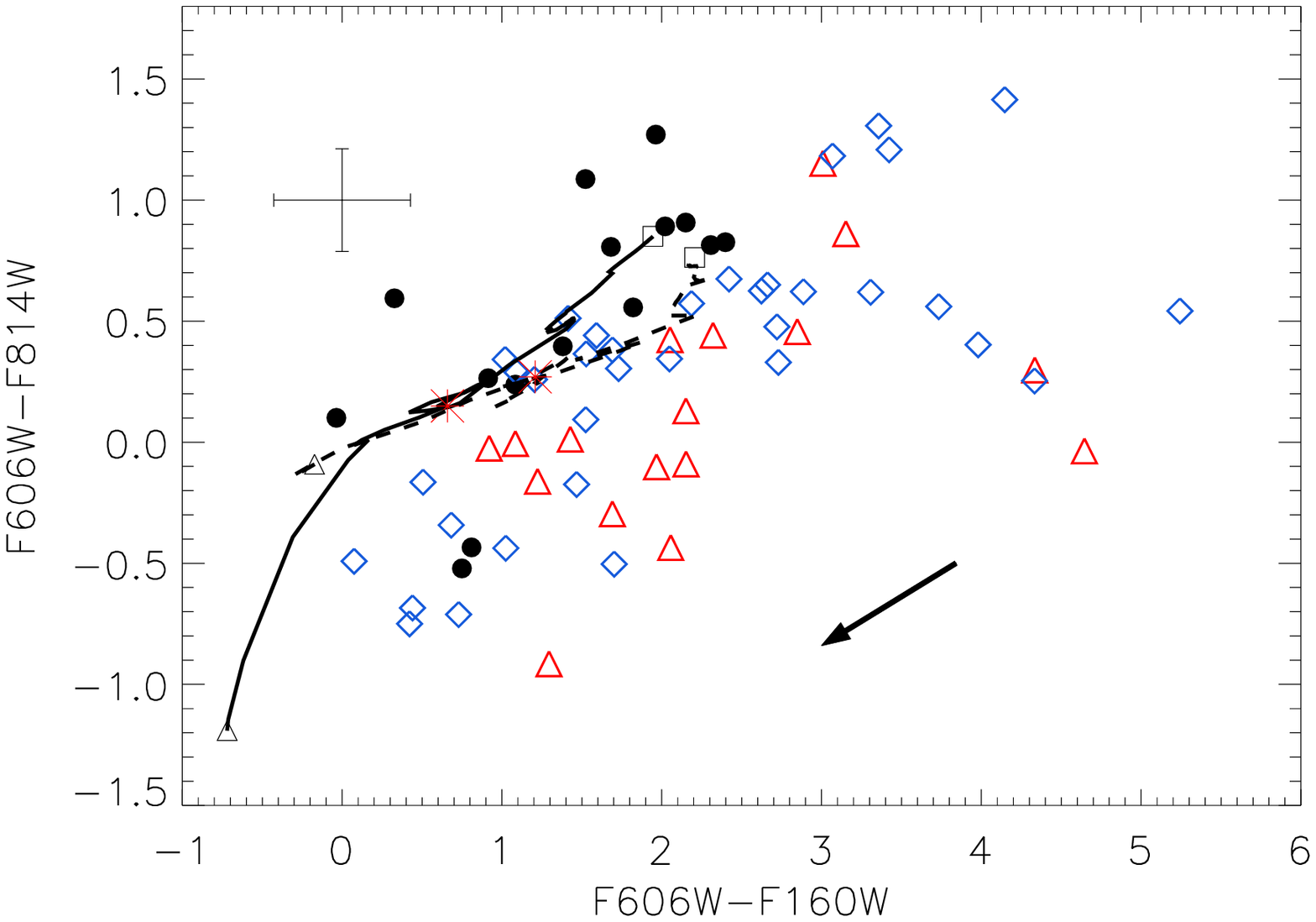}}}
\resizebox{0.45\hsize}{!}{\rotatebox{0}{\includegraphics{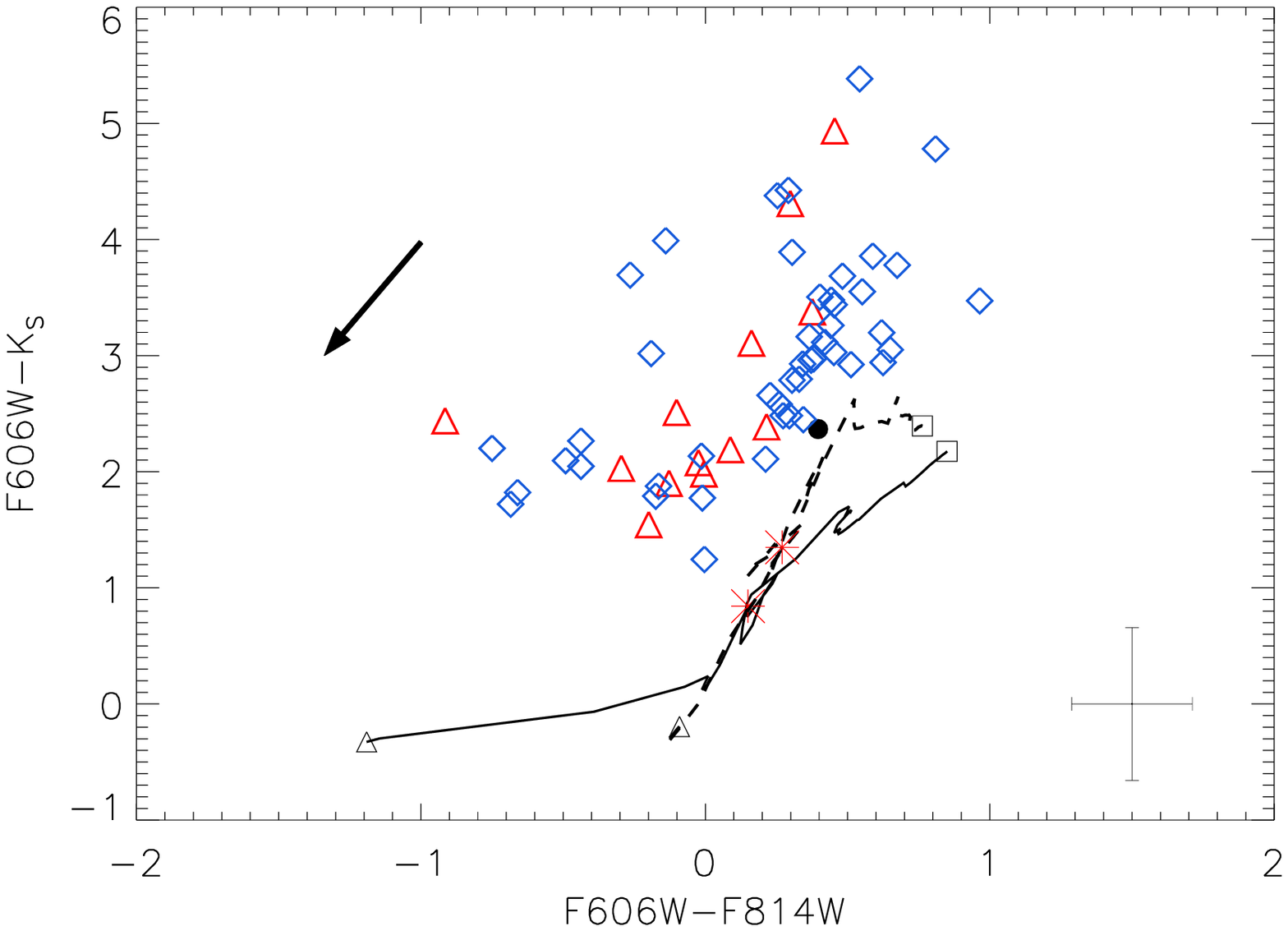}}} 

\caption{Colour-colour diagrams for the cluster samples. The
filled black dots are clusters without a red excess, fitted on
{\it UV-UBVRIHK$_s$}. The open red triangles are clusters with
flux excess in H and K$_s$ and fitted on {\it UV-UBVRI}, and the
blue diamonds denote clusters with excess in all of I, H and K$_s$, fitted only on
{\it UV-UBVR}. The thick black line shows the predicted stellar
evolutionary model according to Z01. The black dashed
line is the corresponding M08 model, with $Z=0.004$. The
black triangles indicate the youngest ages included in each model, 0.55 Myr for
Z01 and 4 Myr for M08.
The red crosses mark the position in the tracks corresponding to 10
Myr and the black squares the end of the tracks at 14 Gyr. In each
plot the mean photometric error associated with each object is
shown. The arrows indicate direction and displacement produced by an
extinction correction of $E(B-V) = 0.3$. The number of objects varies
from plot to plot due to different detection limits in each filter.}
\label{ccd}
\end{figure*}

In Figure~\ref{ccd}, we show four different colour-colour diagrams for
the cluster sample together with the evolutionary tracks by
Z01 for different filter
combinations. M08 evolutionary tracks with the same
metallicity ($Z=0.004$) are also shown. As pointed out in the previous
section, the inclusion of nebular continuum and emission treatment
really do make a fundamental difference in the models at the youngest
ages, and at all the wavelengths (see differences between the
Z01 tracks and those of M08). In
particular the broad filters (F435W, F606W, F814W, F160W) are more
affected, with colours changing drastically in the first 10 Myr (see
F606W$-$F814W, F435W$-$F606W, F160W$-$F606W). Finally, the colour
diagrams which include IR filters also show a difference at older ages
($\sim$ 1 Gyr) between the two evolutionary models which can be
attributed to the better treatment of the AGB-phase by M08.  In the next section we will show, however, that because of
the youth of the clusters in Haro 11, this difference actually only
affects a very small fraction of the sample.

The differences in the number of objects in the colour-colour diagrams
are due to the different detections limits in each
filter. We represent the three cluster subsamples by coloured symbols
corresponding to the flux excess. Black filled dots are clusters which
do not show any flux excess, red open triangles clusters with observed
flux excess in H and K$_s$; blue open diamonds are clusters with an
additional observed displacement in I, and cluster with excess in I and none detections in the IR bands. In the optical and UV colours (the upper
panels of Figure~\ref{ccd}), we see no difference in the distributions
of the clusters with and without flux excesses in I band. In these
panels, many of the clusters can be explained as a young cluster
population affected by different values of extinction.  Few 
clusters, affected by red excess, show very red F606W-F814W colour
($F606W-F814W \geq 1.0$ mag) and are displaced relative to the
tracks. These clusters and many of the blue open diamonds which show
red F606W-F814W colours ($F606W-F814W \geq 0.5$ mag) have UV colours
(left column,  upper row) consistent with
stellar populations younger than 10 Myr ($F140LP-F606W
\sim -2$ mag). 

Including the IR wavebands (Figure~\ref{ccd}, lower panels) in the colors, changes
the situation drastically. Clusters affected by red excess are now
displaced from the model tracks by up to 4 magnitudes. Such offsets
are impossible to reconcile with the estimated values of extinction.
Moreover, several of the sources with red F606W-F814W colour and
strong UV emission, which we expect to be very young, have IR colours
which lie close to the oldest ages of the model tracks at $\sim 14$
Gyr. This is again impossible to explain with the current models.

We analyze possible mechanisms producing the excess at the redder wavebands in section \ref{sec-cause}. In the next section we will focus on the derived physical properties of the clusters.

\subsubsection{Final estimated ages, masses and extinctions}

\begin{figure}
\resizebox{\hsize}{!}{\rotatebox{0}{\includegraphics{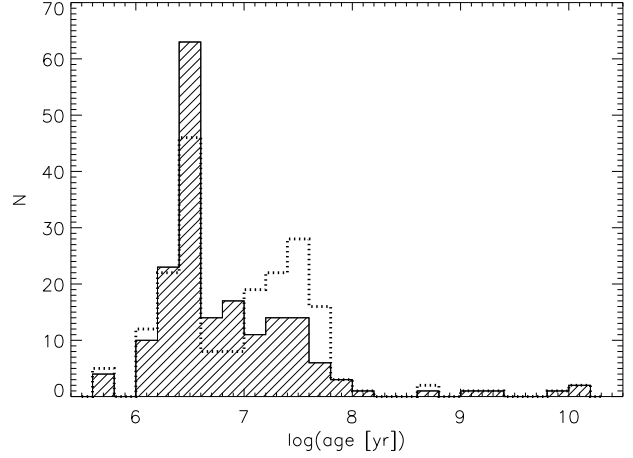}}}
\caption{Final age distribution of the star cluster population (black 
hatched histogram). The dotted over-plotted distribution shows the age
distribution if the SED is fitted over the whole {\it UV-UBVRIHK$_s$}
range, without taking the red excess into account.}
\label{age_distr}
\end{figure}

\begin{figure}
\resizebox{\hsize}{!}{\rotatebox{0}{\includegraphics{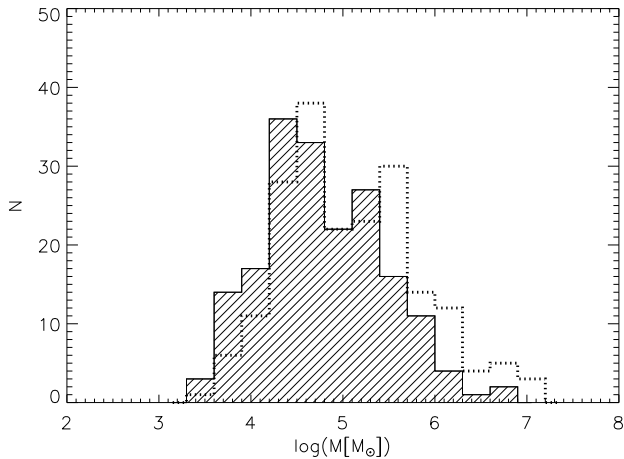}}}
\caption{Final mass distribution (black hatched histogram) of the star 
cluster population. The dotted over-plotted distribution shows the
masses obtained for the clusters if the SED is fitted over the whole
{\it UV-UBVRIHK$_s$} range.}
\label{mass_distr}
\end{figure}

\begin{figure}
\resizebox{\hsize}{!}{\rotatebox{0}{\includegraphics{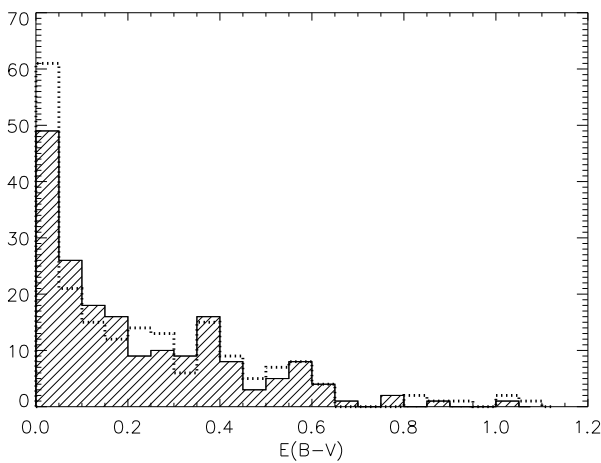}}}
\caption{Final extinction distribution (black hatched histogram) of the star cluster population. The dotted over-plotted distribution shows the
extinctions obtained for the clusters if the SED is fitted over the whole
{\it UV-UBVRIHK$_s$} range.}
\label{ext_distr}
\end{figure}

In Figures~\ref{age_distr}, \ref{mass_distr}, and \ref{ext_distr} we
show our adopted age, mass, and $E(B-V)$ distributions for the total
cluster sample, with 187 objects (black hatched histograms). In this final sample, only  in 21\% of the cases we determined cluster properties using 3 filters, while the 53\% had detection in at least 5 or more bands. The cluster properties of the targets affected by a red excess have been obtained from the fit to the blue-side of the SED spectra.  For comparison we over-plotted the corresponding
distributions (dotted histograms) produced by a {\it UV-UBVRIHK$_s$} fit to the entire sample of clusters. The red excess clearly
introduces a systematic offset in the estimates of all the derived
parameters.  When all the filters are included in the fit, clusters
with a red excess appear to be older and more massive than their
UV-optical photometry alone would indicate. What happens is that when
we attempt to fit the NIR filters affected by flux excess, the best
model choices are old stellar populations in which lower-mass stars
dominate the light. This affects the mass estimation, because the
$M/L$ ratio increases with age: for the same luminosity, older clusters are
also more massive. For example, cluster \# 43 (section \ref{red-excess}
and Figure \ref{spec}) shows that even the excess in I band (at $\sim 8000$
\AA) has important consequences for the estimation of the cluster ages
and masses. 

So far, the filter F814W is one of the most used in the photometric
analysis of the cluster populations in extragalactic environments. Because the
flux excess in this filter is not as dramatic as in the IR filters, it
may go unnoticed when IR data are not available. Indeed, this affected
our own initial analysis of Haro 11's cluster population
(\citealp{b1}). In that paper, the cluster age distribution showed the
same double peak as we have shown here where no correction for red
excess is performed (dotted histogram in
Figure~\ref{age_distr}). Extinction estimates, on the other hand, tend
to be more sensitive to the UV-U fluxes and are therefore not affected
by red excess.

The cluster population in Haro 11 appears to be very young
(Figure~\ref{age_distr}). The present starburst phase started less
than 40 Myr ago with a peak at the extremely young age of 3.5
Myr. Only 16 clusters have estimated ages older then 40 Myr and of
these, half have an age $> 100$ Myr. The age distribution is agreement
with the estimated burst duration, $\tau_{\textnormal{b}} =
M_{\textnormal{burst}}/SFR=35$ Myr, found by \cite{2001A&A...374..800O},
where $M_{\textnormal{burst}}$ is the total mass contained in the burst
and $SFR$. They
estimated an upper limit age for the present burst of 35 Myr, taking the
SFR from the total H$\alpha$ luminosity, uncorrected for
extinction. The age peak at 3.5 Myr is consistent with the young
stellar population (age $< 5$ Myr) which \citet{G2009} identified as
responsible for the hard radiation field inferred from the observed
ratio of [Ne~{\sc iii}]/[Ne~{\sc ii}]$=3.2$ in the galaxy's integrated
mid-IR {\it Spitzer} Infrared Spectrograph spectrum. With about 130 clusters younger than 10 Myr, Haro 11
represents a unique opportunity to investigate the cluster formation
process and evolution at the youngest phase in starburst systems. In
recent reviews, \citet{2009arXiv0911.0796L} and
\citet{2009arXiv0911.0779L} pointed out that clusters younger then
$\sim$ 5 Myr are still partially embedded and affected by several
magnitudes of internal extinction due to the remains of the giant
molecular clouds in which they formed. In Haro 11, the estimated
extinction distribution is quite wide with a peak at $E(B-V) \sim 0.0$
and an extended tail down to at least 0.6 mag
(Figure~\ref{ext_distr}; corresponding to $0.0 \leq A_V
\leq 2.2$ mag). The few cases with $E(B-V)\geq 0.8$ mag have $2.9 < A_V
< 3.7$ mag.  

\begin{figure}
\resizebox{\hsize}{!}{\rotatebox{0}{\includegraphics{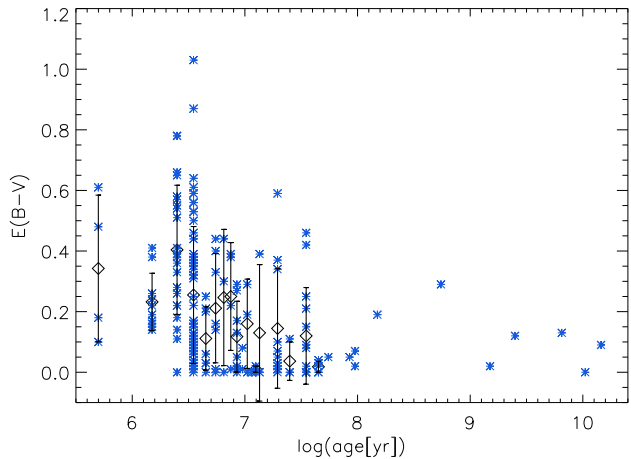}}}
\caption{Extinction versus cluster age. The small blue asterisks are the extinction as function of the age for each cluster. The
black open diamonds indicate the mean value at each age step in the
model, as long as at least three objects are contained in the bin. The
associated error bars are the $1\sigma$ uncertainties.}
\label{ext_age}
\end{figure}

In Figure~\ref{ext_age} we show how the derived
extinction changes as function of cluster age. For ages between 1 and
3.5 Myr the extinction range is wide. Almost all the clusters with
$E(B-V)>0.4$ ($A_V > 1.5$) mag are located in this age range, but the
number of clusters with lower extinction is not negligible. This
observed spread shows the complexity of the cluster formation process
and the gradual dissolution of the parent clouds. A correlation might
be expected between the extinction of clusters younger than 5 Myr and
their masses, but we found no clear trend. This lack of correlation
could indicate that we have underestimated of the mass of partially
embedded clusters. If this is the case, then we should consider the
estimated masses at very young ages as a lower limit to the real
values. At older cluster ages ($> 5 Myr$), we can see in Figure~\ref{ext_age} that
extinction tends to decrease as function of the age, although the
deviations are large. Moreover, previous studies have shown that the
galactic environment is quite complex (\citealp{2002A&A...390..891B},
\citealp{b1}), and the mean extinction in the galaxy is patchy. Some
regions are apparently free from extinction while in others thick
dusty clouds and filaments are apparent (see Figure~\ref{h11}). Finally, we noticed that similar trends between cluster ages and extinctions have also been observed in the Antennae system \citep{2005A&A...443...41M}  and in M51 \citep{2005A&A...431..905B}.

\begin{figure}
\resizebox{\hsize}{!}{\rotatebox{0}{\includegraphics{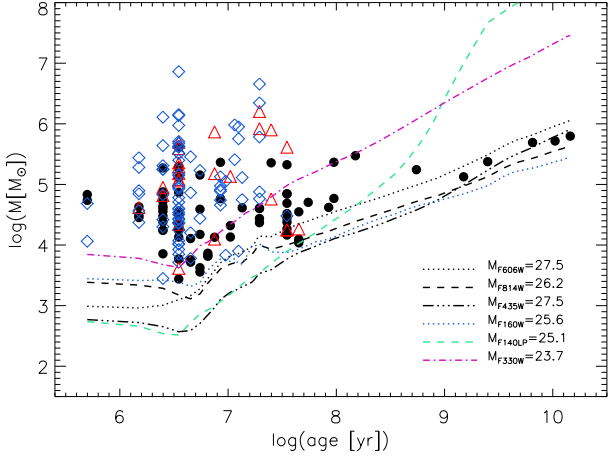}}}
\caption{Ages versus masses for the star cluster population. 
Estimates from {\it UV-UBVRIHK$_s$} fits are shown as filled black
dots, while estimates from {\it UV-UBVRI} fits and {\it UV-UBVR} fits
are shown as red triangles and blue diamonds, respectively. The lines
are the expected detection limits in various filters (see the labels)
according to our stellar evolutionary models.}
\label{massage_distr}
\end{figure}

The clusters are quite massive.  More than half of the targets have
masses between $10^4$ and $10^5 \msun$.  In Haro 11 we have found
around 60 estimated SSCs (clusters with the masses $> 10^5 \msun$),
seven of which have masses larger than $10^6 \msun$. The two knots, B
and C, have masses $\sim 10^7 \msun$. The fit for knot C is poor,
however, and its estimated age, mass and extinction have not been
included in the Figures~\ref{age_distr}, \ref{mass_distr}, and
\ref{ext_distr}. We present a more detailed analysis of these two
regions below. Only $\sim 20$ clusters have masses below $10^4
\msun$, though this small number is affected by the detection limits
we have imposed on the available data. 

In Figure~\ref{massage_distr} we show estimated masses as a function
of our derived cluster ages. The lines indicate the predicted
detection limits using the \citet{Zackrisson et al. a} stellar
evolutionary model, for the corresponding magnitude limits (see Table 1) as
indicated in the plot. Detections below these thresholds are highly unlikely. All
the clusters in the plot have been detected in R and I band (used to
make the initial catalogue; section \ref{photometry}), and at least in
one other filter. While the clusters affected by red excess have a
wide mass range, we noticed that all the most massive clusters ($M
\geq 5\times10^5 \msun$) younger than 40 Myr, appear to be affected 
by at least some flux excess in the NIR. We detected no
clusters older than 40 Myr with any red excess. Six objects with
masses around $10^5 \msun$ and ages greater than 500 Myr have properties
similar to more evolved globular clusters, as in the Milky
Way. 

Curiously, even though restricted detection limits allow us to detect
very young clusters ($1-3$ Myr) with masses around $10^3-10^4 \msun$,
we see almost none in this range of age and mass. This peculiar
phenomenon could be interpreted in different ways. (1) Low mass
clusters $< 10^4$ Myr are still embedded in their parental clouds so
they are not "visible" in optical wavebands. (2) At the distance of
Haro 11, sources may be blended. Indeed, we expect clusters to form in
complexes rather than singly (\citealp{2009arXiv0910.4638E}). If this
is the case, some of the more massive observed clusters may in reality
be clumps of several low-mass clusters. (3) The model, in particular
its handling of the nebular component, has limitations which may lead
to objects at $1-3$ Myr having their ages overestimated. We will
return to these three points in section \ref{CLF}. For the moment, we
note that none of the three explanations excludes the others and all
three could account for the apparent absence of very young low mass
clusters.
\begin{figure}
\resizebox{\hsize}{!}{\rotatebox{0}{\includegraphics{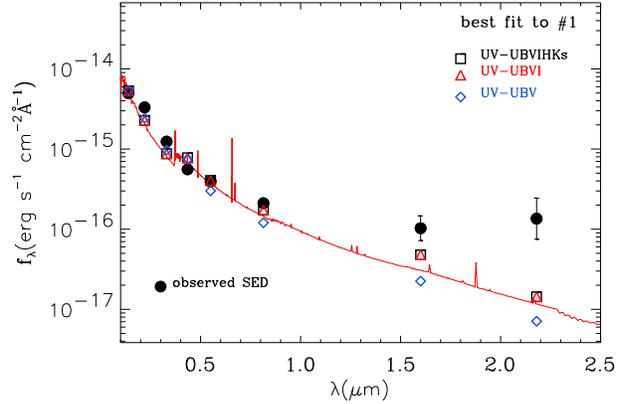}}}
\resizebox{\hsize}{!}{\rotatebox{0}{\includegraphics{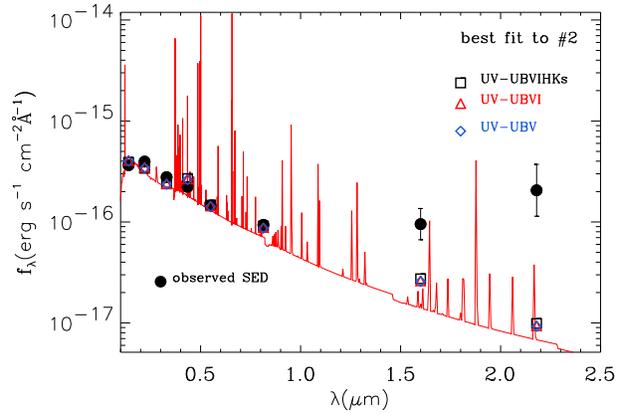}}} \\
\caption{SED analysis of the two knots, C (\#1, above) and B (\#2).}
\label{knots}
\end{figure}

Finally, we have investigated the two most active star forming regions
in the galaxy, knots B and C (Figure~\ref{h11}). We performed aperture
photometry on the two central regions in the same way done for the
point-like clusters even though these two regions appear
extended. Knot C is more reminiscent of a nuclear cluster than
a point-like SC. Dusty filaments appear at the two sides of the
central region in a cone shape. \cite{b1} pointed out that knot C is
the brightest source in the UV, with strong Ly$\alpha$ emission from
its centre, possible indicating that star formation in knot C is still
ongoing.  Knot B
appears obscured by filamentary dust clouds passing through its
central region. It is not clear from our images whether this central
region is composed of a single massive object or several clusters in a
clump similar to the ones observed in the nearby starburst galaxy M\,82
(\citealp{2005ApJ...619..270M};
\citealp{2007ApJ...671..358W}). Knot B is the most massive
cluster-like object after C. Both knots appear affected by the red
excess (see Figure~\ref{knots}) and are very massive objects. The best
fits were obtained with a {\it UV-UBVRI} fit for both clusters. The
physical properties of B and C are summarized in Table
\ref{knots}. As showed by the $\chi^2_r$ value, the fit to knot C is poor so its properties have to be
considered as an approximation. We have obtained spectra of these knots
with VLT/X-shooter which will be the subject of a future paper
(Cumming et al., in prep.). Here we exclude knot C from our final
analysis of the cluster population properties.
\begin{table}
  \caption{Ages, masses, and extinctions for the two star-forming 
regions B and C.}
\centering
  \begin{tabular}{|c|c|c|c|c|c|}
  \hline
  id&$\chi^2_r$&age(Myr)&mass($\msun$)&$E(B-V)$  \\
   \hline
   \hline
  $B$&3.3&3.5&$8.35\times10^6$&0.38\\
  $C$&36.0&9.5&$1.36\times10^7$&0.06\\  
   \hline
\end{tabular}
\end{table}

\subsection{Can a different extinction law give different results?}
\begin{figure}
\resizebox{\hsize}{!}{\rotatebox{0}{\includegraphics{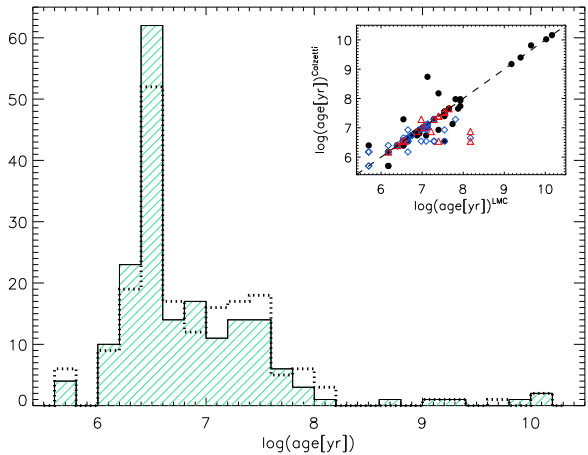}}}
\caption{Recovered age distributions using Calzetti 
(green hatched histogram) and LMC (dotted histogram) extinction
laws. In the inset, we compare for each cluster the age estimated with
the two extinction laws. The different symbols are the same as in Figure \ 11.}
\label{age_comp}
\end{figure}
\begin{figure}
\resizebox{\hsize}{!}{\rotatebox{0}{\includegraphics{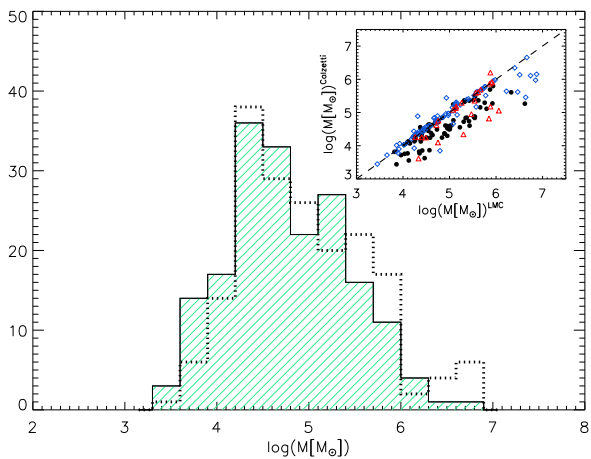}}}
\caption{Recovered mass distributions if a Calzetti (green hatched histogram)
or an LMC (dotted histogram) extinction law is used. The inset shows a
direct comparison is shown between the recovered masses for the two
different fitting runs. The different symbols are the same as in Figure \ 11.}
\label{mass_comp}
\end{figure}
\begin{figure}
\resizebox{\hsize}{!}{\rotatebox{0}{\includegraphics{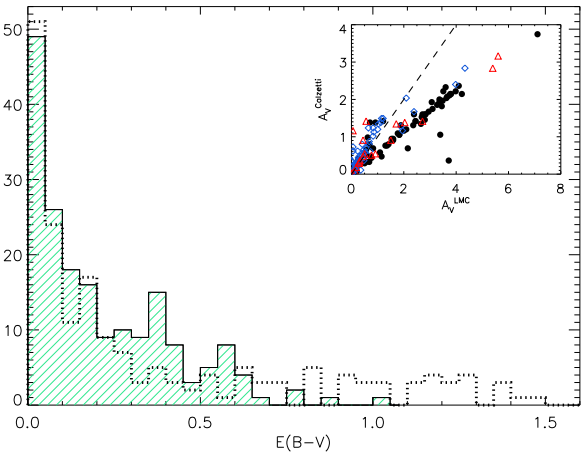}}}
\caption{Recovered extinction distributions using  Calzetti  (green hatched histogram) and LMC (dotted histogram) extinction
laws. The inset shows for each cluster the corresponding visual
extinction, $A_V$ recovered in the two cases. The different symbols are the same as in Figure \ 11.}
\label{ext_comp}
\end{figure}
Could a different extinction law have an impact on our estimates of
the physical properties of the clusters?  To answer this question we
performed a second $\chi^2$ fitting run, this time assuming a Large
Magellanic Cloud (LMC) dust extinction law
(\citealp{1986AJ.....92.1068F}; \citealp{1999ApJ...515..128M}). Both
the LMC and Calzetti \citep{2000ApJ...533..682C} laws have been
obtained from different environments and physical condictions. The Calzetti extinction law \citep{2000ApJ...533..682C}  was derived from the integrated spectra of several starburst galaxies,  so it describes the total extinction affecting a galactic environment and produced by different mechanisms. Since, dust is in general mixed in galaxies, different optical depths are probed as a function of wavelength, and the emerging 'Calzetti law' results from the combination of this effect and the underlying extinction law. On the other hand, the LMC law was determined from the spectra of stars located inside and near the 30 Dor cluster and reproduces the local extinction produced by a molecular dust cloud on single stars, in which case the dust can be accurately treated as a screen. The situtation in Haro11 may, at this distance, be intermediate to these two cases, and most results presented assume the Calzetti law. However it was important to assess whether the LMC law produces different results.  Comparing the outputs of the two runs, we noticed
the following. (1) The values of $\chi^2_r$ obtained with the LMC law are similar to the Calzetti law results and show no improvement in the fit. (2) The clusters still display 
pronounced red excesses. In Figures~\ref{ext_comp}, \ref{mass_comp}, and
\ref{age_comp} we have plotted the recovered extinction, mass, and age
distributions obtained using the Calzetti (green hatched histograms)
and LMC (dotted histograms) laws. The use of a different extinction
law does not have any significant impact on the age distribution of
the clusters. Only a few clusters have discordant ages estimates, as
we show in the inset in Figure~\ref{age_comp}. On the other hand,
about one third of the clusters fitted with an LMC extinction law
appear to have higher estimated masses (Figure~\ref{mass_comp}). This
trend in the estimated mass can be explained by the recovered
extinctions when the LMC extinction law is used. Comparing LMC and
Calzetti $E(B-V)$ distributions (Figure~\ref{ext_comp}), we see that a
considerable number of objects with Calzetti $E(B-V)$ in the range
$0.1-0.6$ appear distributed in a long tail toward higher extinction
values. In particular, the quantitative difference in visual
extinction $A_V$ can differ almost a factor of two, depending of which
extinction law is used (see the inset in Figure~\ref{ext_comp}). This
in turn means that objects affected by higher extinctions have higher
estimated masses. While we cannot draw any definitive conclusion on the choice of the
extinction law, it is clear that a considerable fraction of our
recovered extinctions and masses can depend on the adopted extinction
law. Such uncertainties should be taken into account when cluster
properties are estimated. Age estimates, on the other hand, appear
to be robust.

\subsection{Testing the SED fit results with Monte Carlo simulations}

The outputs recovered by the SED fitting technique are connected to the stellar evolutionary models we used. In order to
estimate the uncertainties in our recovered ages, masses and
extinctions, we used Monte Carlo methods to simulate a fully sampled
population of clusters which resembles the population we observe in
Haro 11. The initial sample contains clusters with three different
mass values ($10^4$, $10^5$ and $10^6 \msun$) for each of 51 age step of the
evolutionary model, and 3 extinctions ($E(B-V)=0.05, 0.3, 0.5$), yielding a total of $3\times 51\times 3=459$ clusters.

For each cluster, we use the spectral synthesis models to calculate its apparent magnitude in all the 9 filters for the mass, age and $E(B-V)$ in question. To these ideal magnitudes that perfectly match the model prediction, we add random photometric errors in accordance with the observed magnitude-error relation in each filter and assuming a gaussian error distribution.  In this way we simulated 1000 realizations of each cluster, the fitted them with our SED tool and compared the input to the output for 1000 hypothetical clusters.
\begin{figure*}
\resizebox{0.7\hsize}{!}{\rotatebox{0}{\includegraphics{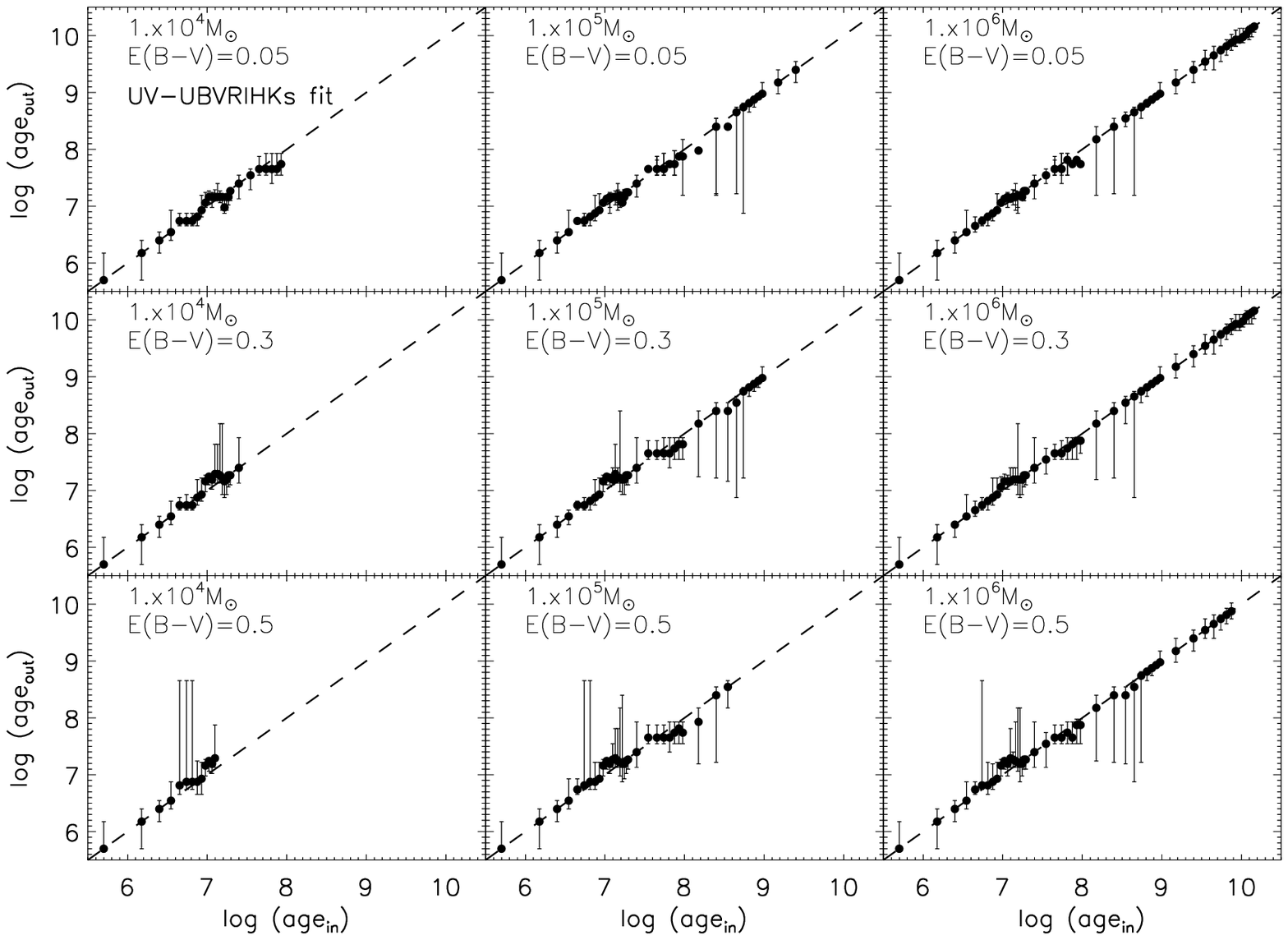}}}
\resizebox{0.7\hsize}{!}{\rotatebox{0}{\includegraphics{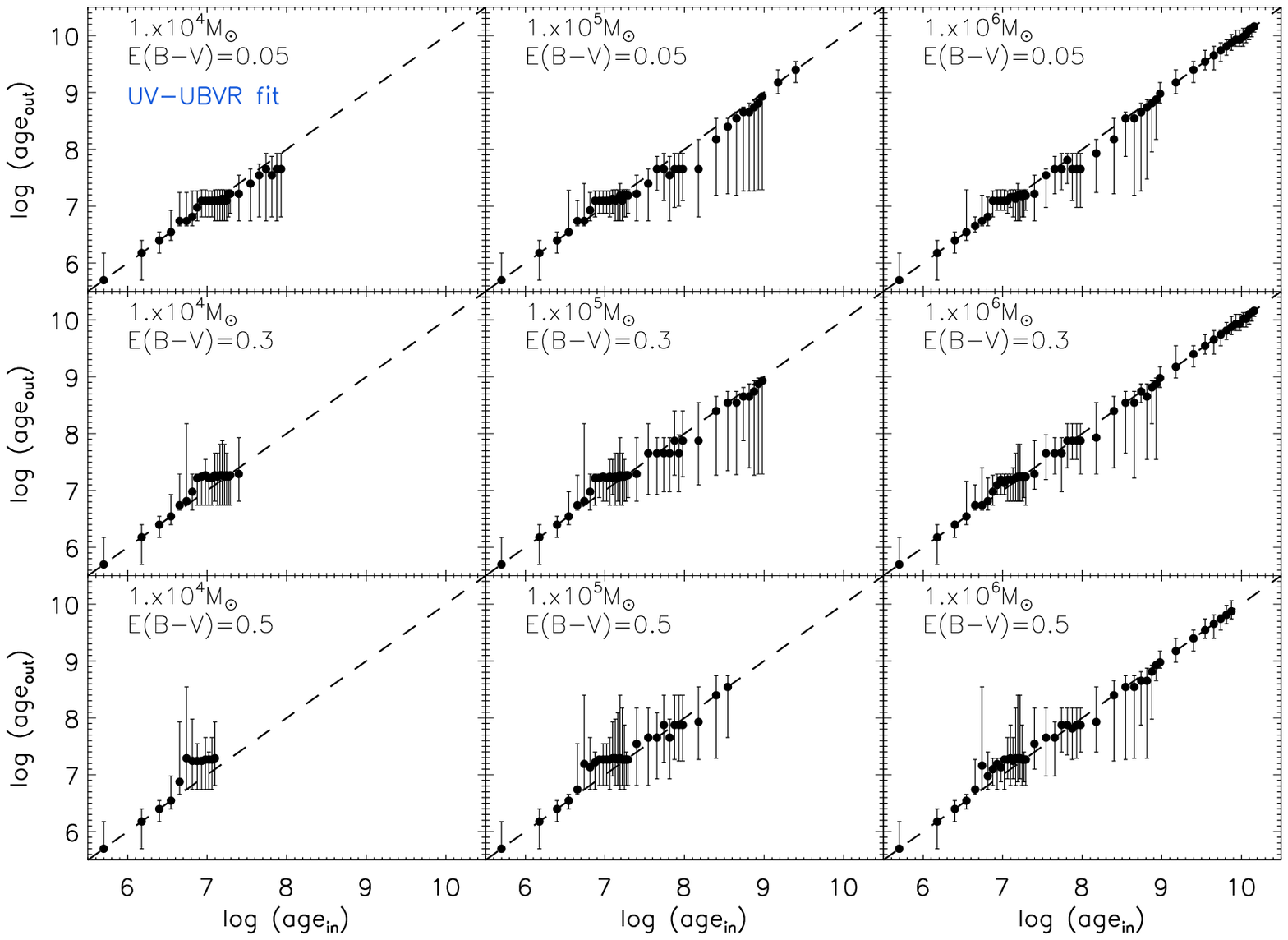}}} 
\caption{Recovered ages as a function of the input ages for different values
of masses and extinctions, as specified in the labels. Nine combinations of mass and extinction are shown for two of the fit ranges  {\it UV-UBVRIHK$_s$} (top), and {\it UV-UBVR} (bottom). The dashed lines
mark where input and output ages are equal. The output ages are the
median values of the 1000 realizations of each cluster and the error bars
are the quartiles of the distribution. Simulated clusters with
apparent magnitude $m_R >27.5$ mag have not been included in the
analysis. This cut in magnitude is reflected in the plots as
incompleteness of the full age sampling.}
\label{age_median}
\end{figure*}

All the 1000 realizations of each cluster were fitted with the same
$\chi^2_r$ fitting program using an {\it UV-UBVRIHK$_s$} fit, an {\it
UV-UBVRI} fit, and a {\it UV-UBVR} fit in turn. The median and
quartiles of the distributions are considered to
be the estimated output age and associated uncertainties,
respectively, for each cluster.  In Figure~\ref{age_median} we show
how the recovered age of an object and its uncertainty changes
depending on the mass and extinction of the object, and the type of fit. From the initial
cluster sample, combinations of age, mass and extinction which
resulted in a cluster with apparent magnitude $m_R > 27.5$ mag was
excluded from the analysis. Looking at the two panels of
Figure~\ref{age_median} we can see that this cut in luminosity appears
as an incomplete sampling at older ages. In general, for the same
mass, clusters with lower extinction are better fitted. The higher the
cluster mass, the smaller the uncertainties at all ages and with
respect to the reddening.  For higher
values of the reddening, input ages of $\sim 10$ Myr tend to lead to
older output ages (and between 100 Myr and 1 Gyr) with a tendency to
underestimate the output ages. This last
effect is less important for higher masses and for lower extinction. 

\begin{figure}
\resizebox{\hsize}{!}{\rotatebox{0}{\includegraphics{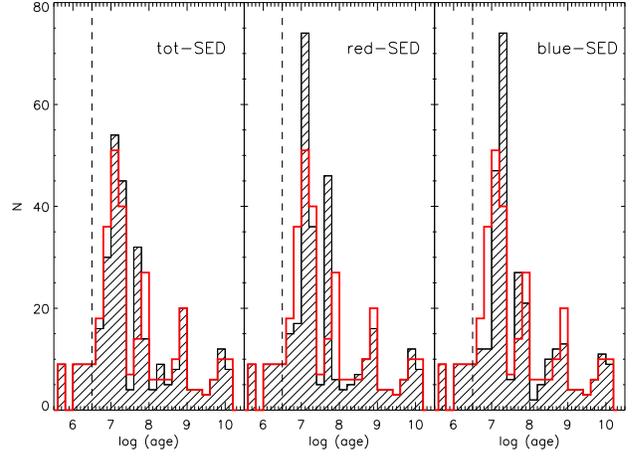}}}
\caption{Recovered age distributions (black hatched histograms) of the 
three sets of fits. From left to right, the recovered age
distributions for {\it UV-UBVRIHK$_s$}, {\it UV-UBVRI} and {\it
UV-UBVR} fits. The red histograms show the input ages of the
simulated sample of clusters.}
\label{age_peak}
\end{figure}

To test for systematic effects of our three different SED fits, we
checked and compared the three sets of recovered ages for the same
mass and extinction distributions. In Figure~\ref{age_median}, we show two of the three SED fits. As expected, if fewer filters are available, the
properties derived from the observed SED become more uncertain, and we see the quartile error bars get wider.  The median
of the distributions, on the other hand, are not affected except at
two critical ages, around 10 Myr and 100 Myr. These age ranges
correspond to particular features of the synthetic stellar population
model. There is a loop in the modeled colors at ages between 10 and 40 Myr which makes it difficult to distinguish
between different ages steps. At intermediate ages, around 100 Myr, the tracks are parallel
to the reddening vector (see Figure~\ref{ccd}, upper left panel) and
clusters could be fitted equally well with younger ages and higher
extinctions. We found no other evidence of systematics
introduced by our analysis. This can be seen also in
the plots in Figure~\ref{age_peak}, in which we compare the input age
distribution (histogram with thick red border) of the simulated
cluster population with the recovered one (black hatched histogram)
derived from the three fit ranges. Once more we exclude clusters with
$m_R > 27.5$ mag from the analysis. When all the filters are included
in the {\it UV-UBVRIHK$_s$} fit, the input and output distributions
differ little. With a {\it UV-UBVRI} fit the disagreement between the
input and output distributions is more evident with a tendency to
overpopulate the age bin around 10 Myr and 60-80 Myr. The exclusion of
the I band in the {\it UV-UBVR} fit moves the peak from 10 to 30
Myr, and the spurious peak a 60-80 Myr remains. In
Figure~\ref{age_distr} we see that the introduction of a subsample of
clusters analyzed with {\it UV-UBVRI} or {\it UV-UBVR} fits produces a
peak in the age distribution at 3.5 Myr ($\log($age$)=6.3$), populated
by objects from older age bins, like 35 Myr. In the distributions of
the simulated cluster sample we used the same age bin size as in
Figure~\ref{age_distr}. None of the fits performed introduce any
displacement to or spurious peak at 3.5 Myr. For this
reason, we are quite confident about the robustness of our estimates.

\section{What is the cause of the red excess?}
\label{sec-cause}
Following our analysis of the cluster photometric and physical properties,
we address the question of what causes the excess of the observed fluxes longward of 8000 \AA  with respect to the synthetic evolutionary models. Clusters with red excesses have also been observed in other
galaxies (\citealp{2002AJ....124.1418W};
\citealp{2005A&A...433..447C}; Reines et al. 2008a, b; \citealp{2009MNRAS.392L..16F}). They are
typically very young and bright at IR wavelengths, and appear to be
subject to by significant dust extinction at optical wavelengths. Their location in colour-colour diagrams tend to be offset
toward both red $V-I$ and
$V-K$ colors. An observed flux excess in I band was found by
\citet{2008NGC4449R} in the young clusters of the galaxy NGC
4449. A subsequent spectroscopic analysis of two of these clusters
\citep{R2009} revealed that the flux excess could be explained if a
contribution from nebular continuum and emission lines were taken into
account. This explanation can be excluded in the
case of the I-band excess in Haro 11's clusters, since our models
already include both nebular continuum and emission lines.

\begin{figure}
\resizebox{\hsize}{!}{\rotatebox{0}{\includegraphics{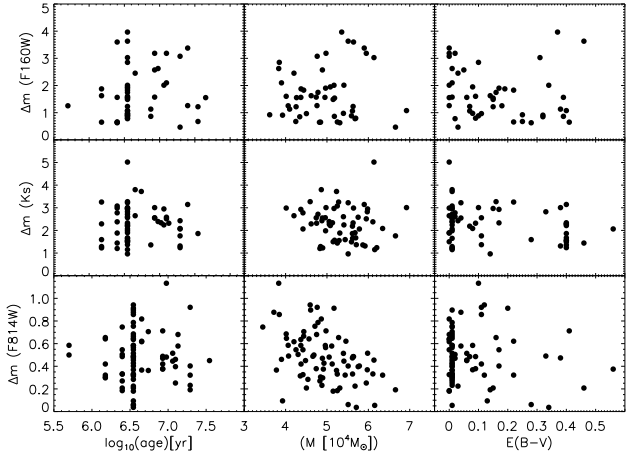}}}
\caption{Residuals of the chi-square fits in I (bottom), K$_s$ 
(middle) and H (top) versus the estimated ages (left), masses (center)
and extinctions (right).}
\label{delta_prop}
\end{figure}

Thanks to the large number of clusters with red excesses in Haro 11,
we were able to check possible correlations between the magnitude of
the excess at $\lambda>8000$\AA\ and the cluster physical properties.
\begin{figure}
\resizebox{\hsize}{!}{\rotatebox{0}{\includegraphics{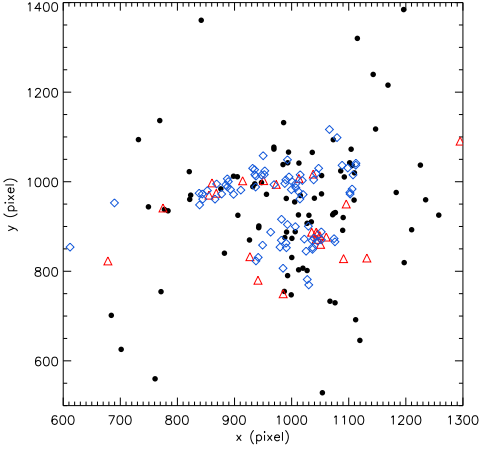}}}
\caption{Position in the galaxy of the three subsamples of clusters. The
normal clusters are labeled as filled black points. Clusters with flux
excess in H and K$_s$ are denoted by red triangles; clusters with
excess also in I band are the blue diamonds.}
\label{re_gal}
\end{figure}
First of all, we looked at the position in the galaxy of clusters
affected by a red excess (Figure~\ref{re_gal}). They are mainly
located in the three starburst knots, as one might expect from their
young ages. Comparing Figure~\ref{h11} and Figure~\ref{re_gal} shows
that that many clusters with flux excesses are located close to dusty regions. Nevertheless, we found no clear relation between the
measured cluster excesses and the corresponding estimated extinctions
(Figure~\ref{delta_prop}, right column).

We have looked for further clues to the red excess by comparing the
deviations from the best SED fits ($\Delta m$) with the derived
properties of age, mass and extinction. In Figure~\ref{delta_prop} we plot the measured flux excess in I, H
and K$_s$ against the recovered ages, masses and extinctions of the
clusters. We see no correlation between the strength of the
displacement and any of these properties. The only correlation with
age, visible in Figure~\ref{massage_distr}, is the absence of clusters
affected by a red excess with ages older than 40 Myr.  This result may
however be biased by the small number of objects in the older age
bins. In summary, the lack of any clear relations seems to indicate
that the red excess could be produced from different mechanisms in different clusters.

As first discussed by \citet{2008NGC4449R}, different mechanisms could cause the red excess at such young ages. For clusters younger than 5 Myr, a possible explanation for the
observed I-band excess is the extended red emission (ERE) phenomenon. ERE
has been observed in a wide range of environments, among them star
forming regions (see \cite{2004ASPC..309..115W} for a review) and it
affects mostly optical and near-IR wavelengths, between
7000-9000 \AA. The effect is expected to be strong in presence of
strong UV radiation fields (as Haro 11 has been shown to possess;
\citealp{G2009}) mainly produced by the most massive stars in the
clusters. However as we have noted, this mechanism cannot explain the
excess in the infrared, so a different physical process needs to be
considered. 

Clusters younger than 5 Myr old are still partially embedded in the original dust
cloud. At this stage the complexity of the cluster environment and the uncertainties in the used models are quite large. Hot
dust emissions can produce a "rise" at IR wavelengths. The
strong UV radiation field produced by the most massive stars in the
cluster is expected to be absorbed by dust grains, which eventually
reemit photons at IR wavelengths depending on the size and temperature
of the grains. Since Haro 11 has a total FIR luminosity
that qualifies it as a LIRG, such processes are undoubtedly working in
the galaxy (\citealp{G2009}); \citealp{2009AJ....138..130S}). Similar explanations were first suggested by \citet{2008sbc0335R} to explain the red excess they observed in the SSCs of SBS 0335-052.   New
fits to those clusters (Adamo et al., to be submitted) show that the
excesses in these clusters can be explained by nebular emission, a
possibility which was not properly addressed by \citet{2008sbc0335R}. On the
other hand, at such young stages, it is possible that the clusters
have two stellar components, one of which has
been able to shed its dust envelope and
contribute to the UV-optical emission while the other is still
embedded in the star-forming cloud and still highly
extinguished. These embedded stars are pre-main sequence objects and young stellar objects
(YSOs) still surrounded by circumstellar disks that can produce this IR-bright component. Modelling is required in order to estimate how
massive should be the second stellar component in order to produce
such large displacements in H and K$_s$ band. Moreover, the presence of an embedded stellar
population contributing only in the IR would imply an incorrect mass
estimation for the clusters.

The UV radiation field is expected to soften after a few Myr as the
most massive stars evolve and finally explode as supernovae. This
means that in clusters older than 5 Myr the I-band excess cannot be
explained by the ERE. We also expect that for ages $> 5$ Myr, the
embedded phase is over and YSOs and pre-main sequence stars have
evolved so that they do not contribute anymore to the IR fluxes.

Two further effects could explain the presence of a NIR excess in
older clusters. (1) Blends of two or more clusters may well be present
in our data. The pixel size at distance of Haro 11 corresponds to $\sim$ 9 pc. If, for example, an older cluster is seen in
the line of sight towards a very young, embedded cluster, the cluster
behind would only be visible at IR wavelengths, while the estimated
age and mass are for the UV-optical bright cluster in front.

(2) Uneven IMF sampling in less massive clusters can also affect the
expected NIR flux.  Studies of the evolution of the stellar isochrones  (\citealp{2009ApJ...699.1938M}, among many others) have
shown that red supergiants (RSGs) dominate the star cluster light in
red and near-IR wavelengths as early as 6 Myr. Z01 models are constructed by integrating the contribution to the
cluster's light from all the stars populating a well sampled IMF,
normalized with respect to the mass of the cluster. This method is, however,
valid only if the stellar population has a total mass of at least
$10^6 \msun$, otherwise uneven sampling introduces important
changes in the colors of the clusters. In
fact, the IMF describes a probability distribution of the mass at
which stars form. This means that, in nature, the IMF sampling is a
purely stochastic event. For clusters with masses not large enough
to fully sample the IMF, the colours of the clusters are dominated
by fluctuations around the expected values (see
\citealp{2004A&A...413..145C}, \citealp{2002A&A...393..167L},
\citealp{2009arXiv0903.4557L} and \citealp{2009arXiv0908.2742F}). At
cluster ages between 6$-$60 Myr, RSGs dominate the light at red and
near-IR wavelengths. If the observed cluster happens to have a larger
number of RSGs than predicted by the model, this would be
seen as an excess in the observed flux at $\lambda > 8000$
\AA. Simulated clusters with
stochastic IMF sampling deviate from the expected evolutionary tracks
by as much as $1.0<V-H<5.0$ ($0.5<V-I<2.0$), compared to the expected
$0<V-H<1$ ($-1.0<V-I<0.3$) colours (M.\ Fouesneau, private
communication).

Finally, a bottom-heavy IMF could in principle explain the IR excess over all the age range. However it is difficult to
imagine why only a fraction of the clusters would manufacture stars
with a different IMF. A recent review by \citet{2010arXiv1001.2965B}, shows observational evidence for a universal IMF shape in different galactic and extragalactic environments.
Finally, the total stellar mass necessary to
produce such an excess would be very high and unphysical for "normal" clusters.
 
\section{Global starburst properties and evolution derived from 
the star clusters}

\begin{figure}
\resizebox{\hsize}{!}{\rotatebox{0}{\includegraphics{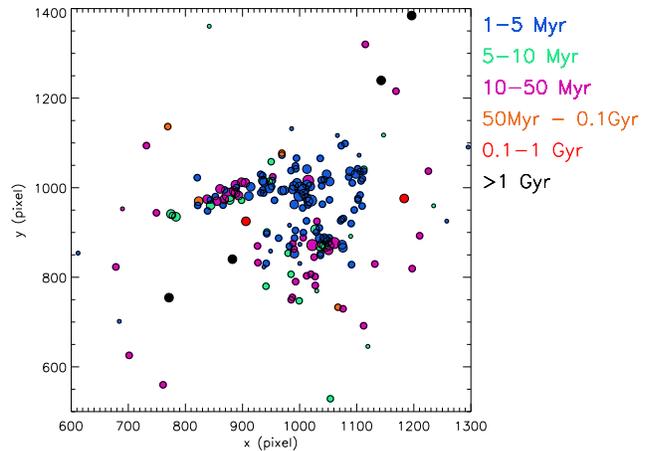}}}
\caption{Spatial distribution of the SCs in Haro 11. The different 
circle sizes correspond to four mass ranges ($M \geq 10^4$, $10^4-10^5$,
$10^6-10^7$ and $>10^7 \msun$). The age ranges are defined by colours as
indicated in the legend.}
\label{scpos}
\end{figure}

In Figure~\ref{scpos} we show the position of the clusters in the
galaxy as function of their estimated ages (colours) and masses
(circle sizes). The youngest clusters (in blue) are located mainly in
close to knot B. The locations suggest a starburst propagating from
the the older regions C and A toward knot B. Older clusters are
located throughout the main burst regions. The clusters outside the
main starburst appear to follow a spiral arm-like distribution, but in the absence of information on
the 3-dimensional distribution we do not explore this possibility
further. 

\subsection{The cluster luminosity function and the fraction of 
missing clusters} \label{CLF} 

\begin{figure}
\resizebox{\hsize}{!}{\rotatebox{0}{\includegraphics{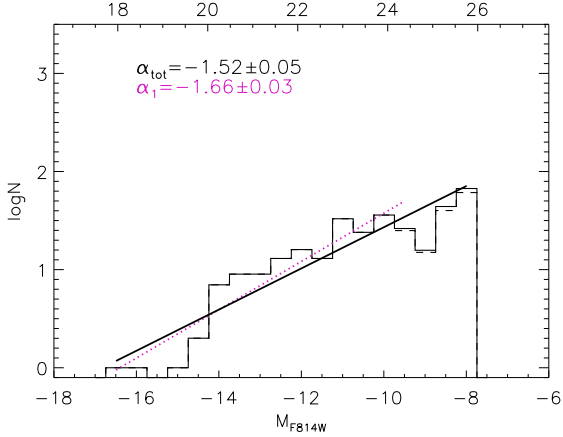}}}
\caption{Observed CLF in the F814W filter with (solid 
line histogram) and without (dashed line histogram) corrections for
incompleteness. The plot shows both absolute and apparent magnitudes. The thick black line shows the best fit to the
completeness corrected distribution; the slope $\alpha_{\rm tot}$ is indicated in
the plot. Finally, the dashed line in magenta is the best fit line to
the distribution with lower magnitude limit set to $25.0$ mag (the 90
\% completeness limit for cluster inside the crowded region; see
section \ref{completeness}). The slope $\alpha_1$ is labelled in
magenta.}
\label{LF}
\end{figure}

In Figure~\ref{massage_distr} we pointed out a deficit of low mass
clusters in the age range of 1-3 Myr. Using our SSP model, we
estimated that clusters at these ages and masses should have 
apparent magnitude in I band in the range $23.0-26.0$ mag. In section
\ref{completeness}, we have showed that in the crowded,  
most active star-forming regions where we expect to find mainly young
clusters, detection limits went below 90\% already at $\sim 25$
mag. Therefore it is possible that a fraction of low mass clusters is
lost either because they are heavily embedded and fall below the
detected thresholds imposed by the available data. Blending may also
affects our sources in that we are not able in some cases to resolve
single clusters. We consider it very unlikely that the missing
clusters have already been destroyed (so-called "infant mortality", \citealp{2003ARA&A..41...57L}). For help with this issue, we decided to investigate
the cluster luminosity function (CLF). We define the CLF as
\begin{equation}\label{alphaeq1}
dN(L_i)=BL_i^\alpha dL_i,
\end{equation}
where $L_i$ is the luminosity of the clusters in a given waveband $i$, and $B$ is a constant. In order to determine the slope, $\alpha$, we fit a
linear relation in $\log-\log$ space,
\begin{equation} \label{alphaeq2}
\log N(L_i) = A M_i + D,
\end{equation}
with the variable $A$ related to $\alpha$ according to
\begin{equation}\label{alphaeq3}
A=-0.4(\alpha+1).
\end{equation}
In our specific case we excluded the R filter (F606W) because it is
heavily contaminated by H$\alpha$ emission at the youngest ages,
yielding a shallower CLF. The F814W (I) band is instead more suitable
for constructing the CLF. As we stated in section \ref{cluster-det},
563 objects were detected in both I and R bands. To avoid
uncertainties due to underestimates of the completeness, we used only
clusters detected above the 90\% completeness limit in I ($< 26.5$
mag), giving a total of 304 objects. In Figure~\ref{LF}
we show the resulting CLF for the I band in absolute magnitudes,
adopting a distance modulus of $M-m=34.58$.
The thick black line is the best fit to the total distribution,
corrected for incompleteness. The obtained slope $\alpha_{\rm tot} =
-1.52 \pm 0.05$, below the typical observed value, $\alpha \sim
-2$. Fitting a double power law produces small difference between the
slope in bright and fainter bins so that does not improve the fit. However we noticed in section \ref{completeness} that in
crowded regions the 90\% completeness limit is reached already at
$\sim 25.0$ mag. We therefore performed a conservative fit including
only objects brighter than 25.0 mag. The recovered slope is steeper
($\alpha_1 = -1.66 \pm 0.03$) but still not close to $-2$. We also
tried constructing an extinction-corrected CLF, using this time only
clusters with known extinction. The recovered slope of the
distribution, $\alpha = -1.64 \pm 0.02$, was still unchanged. 

It is known that crowding and projection effects can flatten the slope
of the LF \citep{2002AJ....124.1393L}. At the distance of Haro 11, and with a
pixel resolution of $\sim 9$ pc in the F814W {\it HST} filter, some of the clusters are resolved. With an aperture radius
of 4 pixels, our photometry was carried out in a region of diameter
$\sim 72$ pc, much larger than typical cluster diameters (5
pc). Bright detections may therefore be the result of overlapping
sources. By substituting, in the equation (3), $\alpha=-2$
and the constant $D$, given by the assumption that the number of
objects in the third bin is correct (we exclude the two brightest
objects, which are knots B and C), we made a rough estimation of the
missing fraction of clusters, using the relation (2). This leads to an estimate of 85\% of objects missing in our
analysis. However, this fraction seems unreasonably high and
impossible to reconcile with the number of detected clusters and the output of our completeness
test. While we are certain that blending affects our data, we cannot
give a quantitative estimate of the effect on the
CLF. Moreover, some of the low mass clusters with ages between $1-3$
Myr are still in the embedded phase and may fall below the optical
detection limits due to extinction. Interestingly, in our group's previously published CLF study of SCs in another BCG, ESO
338-IG04, by \cite{2003A&A...408..887O}, the slope found was $-1.8\pm0.15$. This galaxy lies close enough ($\sim 37.4^3$ Mpc)
for blending between systems not to be an
issue. However, the steepness of the CLF is still below the expected
value. It is possible that the physical environment in BCGs differs
from other late-type galaxies, perhaps related to their status as
merging and interacting systems. It has been already
noticed that a number of dwarf starburst galaxies contain highly
luminous clusters much brighter that expected from the general size of
their cluster population (e.g. NGC 1569, \citealp{2004MNRAS.347...17A}; NGC
1705, \citealp{2002AJ....123.1454B}; NGC 4449, \citealp{2008NGC4449R}; SBC
0335-052, \citealp{2008sbc0335R} ). \cite{2002AJ....124.1393L}
refers to NGC 1569 and NGC 1705 as an example of possibly biased CLF
toward high-mass clusters or in other words that CLF in those galaxies
may be peculiar. If this is true, then BCGs are more similar in this
respect to dwarf starburst galaxies than to very
massive systems.

\subsection{CFR versus SFR in Haro 11}
\label{cfr_sfr}
The star formation rate in Haro 11 is very high. \cite{b1} found a SFR
$\sim 19 - 25 \msun $yr$^{-1}$, using a wide range of estimators. Applying the relation in
\cite{1997MNRAS.289..490R} we calculated a total IR luminosity of
$L_{FIR} = 4.43 \times 10^{44}$ erg~s$^{-1}$ from the IR luminosity at
60 $\mu$m (taken from NED).  Using the total IR
luminosity of Haro 11 we derived the present SFR from the Kennicutt's
relation \citep{1998ARA&A..36..189K} of SFR $= 21.1 \msun$ yr$^{-1}$,
in perfect agreement with the estimation made by \cite{b1}. In the
present work we used a mean value of the SFR $= 22 \pm 3 \msun
yr^{-1}$.

Typically, a fraction of the star formation in the galaxy occurs in
bound clusters, which will possibly survive and
evolve. The remaining star formation occurs in low-mass clusters
(analogous to the Milky Way's open clusters), associations and more
extended systems which would be destroyed during the first few Myr and
dispersed into the field, as predicted by infant
mortality. \cite{2008MNRAS.390..759B} has defined the present cluster
formation efficiency in a host galaxy as $\Gamma = $CFR/SFR, where
CFR$=M_{tot}/\Delta t$ is the present cluster formation rate at which
the galaxy produces a total cluster mass $M_{tot}$ in a given age
interval $\Delta t$. In a recent paper, \citet{2010arXiv1002.2894G}
have found a correlation between the value of $\Gamma$ and the star
formation density $\Sigma_{\textnormal{SFR}}$ of the hosting system,
confirming that higher SFR leads to the formation of more bound
clusters.
\begin{figure}
\resizebox{\hsize}{!}{\rotatebox{0}{\includegraphics{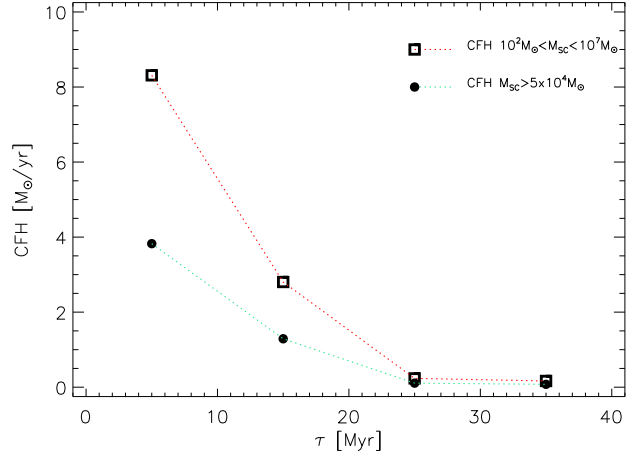}}}
\caption{Cluster formation history (CFH) of the present starburst in Haro 11. The 
filled black dots connected by the green dotted line shows the
observed CFH estimated by the total mass in clusters more massive than
$5\times10^4 \msun$. The open squares connected by the dotted
red line trace the extrapolated CFH for clusters with masses in the
range $10^2$ to $10^7 \msun$, under the condition of an ICMF with index
$-2$ and assuming that we know the total mass in clusters more massive
than $5\times10^4 \msun$ in each age bin.}
\label{CFH}
\end{figure}

The cluster age distribution for Haro 11 shows that the present
starburst has lasted for 40 Myr, with a peak of cluster formation only
a few Myr ago. Under the condition that a full cluster
population forms in 10 Myr, we tried to
estimate how the rate of cluster formation has changed since the burst
phase began. Taking into account the distance of the galaxy and
possible blending at the faint end of the CMF, we estimated the total
mass in clusters more massive than $5\times10^4 \msun$ at each 10 Myr
age bin, assuming that we detect all the clusters younger than 40 Myr
with masses above this limit. In this way we have constructed the
observed cluster formation history (CFH) in clusters more massive than
$5\times10^4 \msun$ (Figure~\ref{CFH}). The CFH of massive clusters
peaks with the youngest cluster population, which has formed in the
last 10 Myr at a rate of 3.8 $\msun
$yr$^{-1}$. This shows that at the present time, $\Gamma
(M_{\textnormal{SC}} > 5\times10^4 \msun) \sim 0.17 $, or in other
words, 17 \% of the star formation is happening in massive
clusters. The CFR declines toward lower rates at larger lookback times
and reaches a value of $0.08 \msun $yr$^{-1}$ when the cluster
population formed around 40 Myr ago. This decline may also indicate
lower star formation rates at this epoch. However the
massive cluster formation in the galaxy during the whole
period has been extremely efficient.
\begin{figure}
\resizebox{\hsize}{!}{\rotatebox{0}{\includegraphics{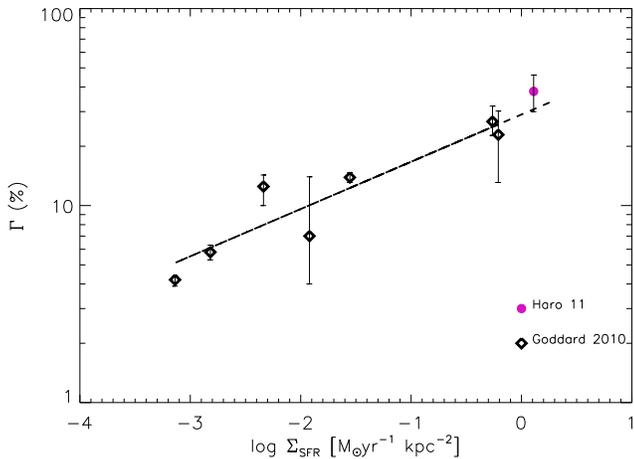}}}
\caption{Cluster formation efficiency, $\Gamma$, as function of the
galactic star formation rate density, $\Sigma_{SFR}$. The black diamonds
are the galaxy sample of \citet{2010arXiv1002.2894G} which were used
to obtain the best fit power-low relation shown by the dashed line
(\citealp{2010arXiv1002.2894G}, their equation (3)). At the right hand
end we show the position of Haro 11 (dot) which fits the relation nicely
despite its extreme $\Gamma$ and SFR values.}
\label{CFR}
\end{figure}

In order to constrain a total cluster formation rate, we estimated the
missing fraction of mass in low mass clusters
($10^2<M_{\textnormal{SC}}<5\times10^4 \msun$), taking an initial cluster mass function (ICMF) with
slope $-2$ and assuming that the total cluster mass in the range $5\times10^4 -
10^7 \msun$ is known. The extracted total CFH is also shown in
Figure~\ref{CFH}. The total CFR at ages below 10 Myr is $8.31
\msun$yr$^{-1}$, which means cluster formation efficiency $\Gamma
\sim 0.38 \pm 0.6 $ or $\sim 38$ \%. This fraction is extremely high
compared to the value observed in cluster systems in other galaxies
(see Table \ref{SFRs}), and almost a factor of 10 higher than in the
SMC ($\Gamma = 2 - 4 \%$; \citealp{2008A&A...482..165G}) and the
solar neighborhood ($\sim 5\%$; \citealp{2003ARA&A..41...57L}). We reach the
same result if we estimate the present CFR from the most
massive cluster, $M_{max}$, in the age bin 1-10 Myr. In fact
\cite{2008A&A...482..165G} showed that $M_{\rm tot} \simeq 10 M_{max}$,
for a constant cluster formation rate and a power-law CIMF with index
$-2$, and assuming that the maximum cluster mass formed by the galaxy,
$M_{\rm up}$ is bigger than the $M_{\rm max}$ observed in the
system. If we assume that the CFR is constant over the last 10 Myr and
that the most massive cluster formed\footnote{The very massive region
$C$ is excluded due to the uncertainties in its SED fit.} is knot $B$
with a mass of $8.35 \times 10^6 \msun$, the present CFR is $8.35
\msun yr^{-1}$, very similar to the value estimated above.
\begin{table*}
\caption{The cluster population of Haro 11 compared to the galaxy sample of \citet{2010arXiv1002.2894G}. We show the SFR, $\Gamma$(\%) and total mass in
clusters younger than 10 Myr and more massive than 10$^3 \msun$ (NGC
1569, M83, NGC 6946) or $5\times10^4 \msun$ (NGC 3256 and Haro 11),
M$_{\textnormal{YSC}}$, the total hosting galaxy mass,
M$_{\textnormal{gal}}$, and the fraction of the total cluster mass
with respect to the galactic mass,
$f=$M$_{\textnormal{YSC}}$/M$_{\textnormal{gal}}$. The data of the
galaxies NGC 1569, M83, NGC 6946, and NGC 3256 are from
\citet{2010arXiv1002.2894G} unless otherwise specified. (a)
\citet{1988A&A...194...24I}; (b) \citet{2004A&A...422..865L}; (c)
\citet{2002A&A...388....7W}; (d) \citet{1978MNRAS.185...31F}. }
\centering
  \begin{tabular}{|c|c|c|c|c|c|}
  \hline
 galaxy& SFR($\msun$ yr$^{-1}$)& $\Gamma$(\%)& M$_{\textnormal{YSC}} (\msun)$&M$_{\textnormal{gal}} (\msun)$ &$f$ \\
   \hline
   \hline
   NGC 1569&0.3626&13.9$\pm$0.8&3.52$\times10^5$ &4.6$\times10^8$(a)&0.0008\\
  
    M83 &0.3867&26.7$\pm^{5.3}_{4.0}$&7.24$\times10^5$ &6.1$\times10^{10}$(b)&1.5$\times10^{-5}$\\
  
    NGC 6946&0.1725&12.5$\pm^{1.8}_{2.5}$&1.284$\times10^5$ &5$\times10^{10}$(c)&2.6$\times10^{-6}$\\
  
    NGC 3256&46.17&22.9$\pm^{7.3}_{9.8}$&1.66$\times10^7$ &$\sim 5\times10^{10}$(d)&0.0003\\
 
   Haro 11&$22\pm3$&38$\pm$6&3.82$\times10^7$ &$\sim 1\times10^{10}$&0.004\\
   \hline
\end{tabular}
\label{SFRs}
\end{table*}

If blending affects our cluster mass estimates then we may be
overestimating the CFR. Applying the relation found by
\citet{2010arXiv1002.2894G} between $\Gamma$(\%) and
$\Sigma_{\textnormal{SFR}}$, we can estimate the expected value of
$\Gamma$ for the SFR observed in Haro 11. The starburst in this galaxy
covers an area of $\sim$2.34 kpc$^2$ so that the
$\Sigma_{\textnormal{SFR}} = 1.28 \msun$~yr$^{-1}$~kpc$^{-2}$. From
equation 3 in \citet{2010arXiv1002.2894G} we find that $\Gamma =
31.0\pm 5$\% which is in quite good agreement with the previously
found value. In Figure~\ref{CFR} we show the relation presented by
\citet{2010arXiv1002.2894G} and the galaxy sample they used to derive
it, with the addition of the position of Haro 11. We observed that in spite of the extreme cluster formation environment in Haro 11, the correlation found by \citet{2010arXiv1002.2894G} between $\Gamma$ and density of SFR is still valid, supporting the theory of a close connection between the global properties of the host galaxy and its star cluster population.

In Table \ref{SFRs}
we summarise the properties of the Goddard et al.\ sample together
with Haro 11. For some of the objects the authors gave the total mass
contained in clusters younger than 10 Myr. We used these values to
estimate at which fraction of the total mass of the system they do
correspond, as indicated in the last column of Table \ref{SFRs}. For
Haro 11 we used only the total observed mass in clusters more massive
than $5\times10^4 \msun$. The comparison between the values for
different galaxies nevertheless gives an idea of the strength of the
cluster formation process in Haro 11.

\subsection{Discussion}

In a recent work, \cite{2009AJ....138.1203M} studied the properties of
clumpy star-forming regions associated with bound clusters in nearby
dwarf starburst galaxies. They found that, independently of their
environmental conditions, the formation of those clumps is given by
random sampling of a mass function of the form $dN/dM \propto
M^{-2}$. We can think the mass function as a statistical relation which is
stochastically realized by star formation
in clusters. The galactic environment limits the range of the stochastic action. For
example, the galactic environmental conditions can limit the maximum
possible mass at which a cluster can be formed
\citep{2006A&A...450..129G}. The galactic environment may also lead to
a characteristic mass, $M_c$, above which a CIMF shaped like a
Schechter function declines exponentially (\citealp{2009A&A...494..539L}; \citealp{2009MNRAS.394.2113G}). In
particular, \cite{2009A&A...494..539L} noticed that extreme
star-forming environments show no characteristic truncation at $10^5 <
M_c < 10^6$ because such galaxies are actually able to form very
massive clusters ($M > 10^6$).  \cite{2009A&A...494..539L} presented two hypotheses regarding the
physical mechanism responsible for the formation of massive
clusters. In the first, such clusters are formed by super-GMCs, with masses of  at least $10^7 - 10^8 \msun$. The
second requires a higher star formation efficiency during the collapse
of the GMC, which in turn implies  compression of the gas to
higher densities. In the Milky Way it has been observed that GMCs have
low densities and form clusters with low star formation efficiency
\citep{2003ARA&A..41...57L} producing low mass clusters.  
In quiescent massive regular spirals like the Milky Way the shear due
to the spiral patterns would then favours the fragmentation of GMCs,
preventing very massive clusters from forming. On the other hand, as
pointed out by \cite{2002AJ....123.1454B}, in dwarf starburst systems
the gravitational instabilities acting on the GMCs are instead much
stronger than the shear from the irregular intergalactic medium,
allowing the observed massive clusters to form.

In Haro 11 we have found many massive clusters, over $35\%$ of which
have masses above $10^5 \msun$. In the previous section we showed that
the efficiency with which the galaxy has formed clusters, in particular
massive clusters, is higher than in merging
systems like NGC 3256. Haro 11 has an irregular morphology and is less
massive than normal spiral galaxies. It seems likely that a previous
merger with a gas-rich system has favoured the inflow and compression
of gas, triggering the formation of the massive clusters. In this way,
the galaxy, despite its peculiarities, easily fits into the wider
picture of a sequence in star formation intensity 
stretching from quiescent to very active modes according to Goddard's
relation. The age distribution of the clusters confirms the present
estimated starburst age of $\sim 35$ Myr found by
\citet{2001A&A...374..800O}. The starburst has propagated rapidly
through the entire galaxy as showed in Figure~\ref{scpos}, consuming
the available fuel and is possibly now settling into a period of more modest activity, as attested by the low H~{\sc i} mass content in the galaxy
\citealp{2000A&A...359...41B}. From the available H~{\sc i} mass, the
remaining life time of the present burst phase is $\sim$ 5 Myr.

A final important issue is raised by the similarity between Haro 11
and LBGs observed at high redshift \citep{2008ApJ...677...37O}.
Nearby Lyman break analogues (LBAs) are attractive objects for
studying star formation modes and morphology in detail which is
impossible at high redshift. \citet{2009ApJ...706..203O} have for
example shown that star formation in LBAs is both very active and
concentrated in clumpy regions which may be composed of 
 supermassive star clusters. In Haro 11, we found a CFR of $38$\%,
somewhat higher than locally observed. This result suggests that the CFRs in
LBGs could be almost a factor of 10 higher than in the Local Group,
implying that at high redshift galaxies formed more numerous and more
massive clusters than we observe in the local Universe. However
investigation of other systems similar to Haro 11 are needed before we
can draw firm conclusions.

\section{Conclusions}

Thanks to the deep {\it HST} data stretching from UV to IR wavebands,
we have carried out a photometric study of the clusters in the
starburst regions of Haro 11. We have constrained the physical
properties of 185 clusters through the analysis of their integrated
SEDs. We summarize the main results here.
\begin{enumerate}
\item More than 60 \% of the star clusters in Haro 11 are younger 
than 10 Myr old. The present starburst phase started less than 40 Myr
ago with a peak as recently as 3.5 Myr. Only 16
clusters have estimated ages greater than 40 Myr and, of these, half
are older than $> 100$ Myr. Haro 11 therefore provides a unique
opportunity to investigate the cluster formation and evolution in a
starburst system in its earliest phases.  For ages between 1 and 3.5
Myr the extinction range is quite wide, consistent with a complex cluster formation process and the gradual dissolution of the
parent clouds. 
\item The clusters are quite massive, with masses between 
$10^3 - 10^6 \msun$. There are around 60 estimated SSCs, 7 of them
with masses larger than $10^6 \msun$, and two very massive knots (B
and C) with estimated masses $\geq 10^7 \msun$ which we could not
resolve into smaller clusters. Very young clusters (1-3 Myr) with
masses around $10^3-10^4 \msun$ are absent from our data even though our detection limits should have allowed their
detection. These missing clusters could still be
embedded in their parental clouds, invisible in optical wavebands, or
blended with other clusters in the most crowded regions.
\item We have discovered that a fraction of the clusters exhibits
a flux excess compared to plausible SED models at
wavelengths $>8000$ \AA.  We have noted that fitting those cluster SEDs
affected by the red excess with normal SSP models leads to overestimation
of the clusters ages and masses. 
In several cases, we have obtained acceptable fits  only at bands blueward of I,  highlighting the potential impact of the 
phenomenon  on  estimations of the cluster properties. Of the  clusters affected by the red excess, none are older than 40 Myr. 
\item We have discussed possible causes for the red excess. 
If the age of the cluster is the key property, then the red excess in
clusters younger than 5 Myr could be due to the simultaneous presence
of very young UV-bright massive stars and the dust left from the
GMCs. The ERE mechanism could then originate the excess at I waveband
while hot dust emission produces a "rise" at IR wavelengths, depending
on the size and temperature of the dust grains. Another possible
mechanism at such young stages could be the presence in the cluster of
two stellar components, one of which has shed its dust envelope,
contributing mainly in the UV-optical wavelengths, while the other
component is still deeply embedded in its star-forming cloud and
highly extinguished. Pre-main sequence objects and young stellar
objects (YSOs) still surrounded by circumstellar disks can produce this IR-bright component. Further modelling will be
required in order to estimate how massive the second stellar component
needs to be in order to produce such large displacements in the H and
K$_s$ bands.  The presence of such embedded stellar populations
emitting only in the IR would also imply that the mass of these
clusters will be underestimated. At ages $> 5 Myr$ we
expect that the embedded phase is over; we have suggested two further
mechanisms to explain the NIR excess for ages older than this. First,
evolved clusters may be observed in the line of sight towards younger,
embedded ones. In such cases, the cluster behind would only be visible
at IR-bands, while the age and mass estimated form UV and optical
bands correspond to the cluster in front. Second, red supergiants,
which dominate the light at red and near-IR wavelengths, could
"redden" the observed colours of the clusters at ages as early as 6
Myr. 
\item For the first time, after an early attempt by 
\cite{2003A&A...408..887O}, we have analysed the cluster luminosity 
function in a BCG. Haro 11's CLF is shallower than observed in other
galaxies. Contamination by blending could have this effect on
the CLF, but we didn't find any evidence of overestimation of the CFR, using the more massive SSCs. Obviously, at the distance of the galaxy and with the present
available data, we cannot avoid crowding and projection
contaminations.
\item More than the 30\% of the star formation in Haro 11 has 
happened in bound clusters which can survive and evolve along with the galaxy. This fraction is higher than observed in
the Milky Way and other merging galaxies like NGC 3256, leading to the
conclusion that the environmental conditions in this BCG have favoured
the formation of massive super star clusters. Due to the similarities
between Haro 11 and the high-redshift LBGs, we suggest that cluster
formation rates at higher redshift may also be enhanced.
\end{enumerate}

\section*{Acknowledgments}
We thank {\it HST-helpdesk} for helpful suggestion. Mark Gieles and Morgan Fouesneau are thanked for the very useful discussions and comments made on this work. A.A. and G.\"O acknowledge the support from the Swedish Research council. G.\"O. is a Royal Swedish Academy of Sciences research fellow, supported from a grant from the Knut and Alice Wallenberg foundation. G.\"O. also acknowledges support from the Swedish National Space Board. E.Z. acknowledges a research grant from the Swedish Royal Academy of Sciences. M.H. acknowledges the support of the Swiss National Science Foundation. We thank the anonymous referee for useful suggestions and discussions.

\appendix
\begin{table*}
\caption{Photometric properties of the 4 clusters showed in Figure~\ref{spec}. Magnitudes are in Vega system throughput.}
\centering
  \begin{tabular}{|c|c|c|c|c|c|c|c|c|c|}
   \hline
ID& F606W &F814W&F140LP& F220W&F330W&F435W&F550M& F160W&Ks\\  
\hline
  \hline 
6&22.97$\pm$0.1   & 22.67$\pm$0.07  &  20.75$\pm$0.06 &   21.25$\pm$0.07  &  22.10$\pm$0.08 &  23.12$\pm$0.05 &   23.21$\pm$0.06 &   21.24$\pm$0.42   & 20.18$\pm$0.66\\
43&21.49$\pm$0.11  &  20.81$\pm$0.05  &  21.70$\pm$0.11 &   21.73$\pm$0.13  &  21.59$\pm$0.09  & 22.54$\pm$0.08   & 22.33$\pm$0.07 &   19.07$\pm$0.4  &  17.71$\pm$0.64\\
87&22.32$\pm$0.19   & 22.16$\pm$0.17 &   19.84$\pm$0.12 &   20.51$\pm$0.14  &   $-$  &22.60$\pm$0.15  &  22.42$\pm$0.13   &  $-$  &  19.21$\pm$0.66\\
113&22.49$\pm$ 0.1  &  22.09$\pm$0.05  &    $-$  &    $-$  &   $-$    & 23.29$\pm$0.06  &  22.85$\pm$ 0.06  &  21.11$\pm$0.42  &  20.12$\pm$0.66\\
  \hline
  \end{tabular}
      \end{table*}  
\section{Check on the calibration of the F160W frame}
We performed 2 different tests on the HST IR-frame in order to verify the goodness of the calibration. First, we compared photometry output of few bright sources between the final reduced frame and one single raw frame. At the same aperture radius, sources on the raw frame showed to be $\sim 0.15 $ mag brighter than the final science frame. However this scatter can be explained by sky background and bias effects corrections performed on the final frame. In the second test, we used an available HST archive {\it NIC2}/F160W exposure of the MRK 930 target. Because of the analogy between the {\it NIC3} data set of Haro 11 and MRK 930 we decided to compare the photometry outputs of MRK 930 {\it NIC2} and {\it NIC3} F160W frames. The {\it NIC2} frames were reduced with {\tt MULTIDRIZZLE}. For small aperture radii, the photometry of sources in the {\it NIC2} frame were brighter than the {\it NIC3} ones with a similar constant scatter. At larger radii, the scatter became rapidly negligible. As expected this behavior is due to the better pixel resolution scale and PSF sampling of the {\it NIC2} camera. However because the photometry converged at the same values this shows the correctness of the calibration, showing at the same time the importance of an accurate aperture correction to infinity.

\section{}

\bsp

\label{lastpage}

\end{document}